\documentclass[useAMS,usenatbib]{mn2e}

\usepackage{tabularx}
\usepackage{graphicx}
\usepackage{txfonts}
\usepackage{longtable}
\usepackage{hhline}
\usepackage{arydshln}
\usepackage{multirow}
\usepackage{lscape}
\usepackage{array}
\usepackage{rotating}

\newcommand{\hersc}{{\it Herschel}}
\newcommand{\lab}{LABOCA}

\newcommand{\spitz}{{\it  Spitzer}}

\newcommand{\msun}{$M_\odot$}
\newcommand{\zsun}{$Z_\odot$}
\newcommand{\mic}{$\mu$m}

\newcolumntype{R}[1]{>{\raggedleft\arraybackslash }b{#1}}
\newcolumntype{L}[1]{>{\raggedright\arraybackslash }b{#1}}
\newcolumntype{C}[1]{>{\centering\arraybackslash }b{#1}}

\newlength{\pointwidth}
\begin{document}

  \title[The thermal dust emission in N158-N159-N160]{The thermal dust emission in N158-N159-N160 (LMC) star forming complex mapped by Spitzer, Herschel and LABOCA}

\author[Galametz et al.]
{\parbox{\textwidth}{M. Galametz$^{1}$\thanks{e-mail: mgalamet@ast.cam.ac.uk}, 
S. Hony$^{2}$, 
F. Galliano$^{2}$,
S. C. Madden$^{2}$,
M. Albrecht$^{3}$,
C. Bot$^{4}$,
D. Cormier$^{2}$,
C. Engelbracht$^{5,6}$,
Y. Fukui$^{7}$,
F. P. Israel$^{8}$,
A. Kawamura$^{9}$,
V. Lebouteiller$^{2}$,
A. Li$^{10}$,
M. Meixner$^{11,12}$,
K. Misselt$^{5}$,
E. Montiel$^{5}$,
K. Okumura$^{2}$,
P. Panuzzo$^{2}$,
J. Roman-Duval$^{11}$,
M. Rubio$^{12}$,
M. Sauvage$^{2}$,
J. P. Seale$^{11}$,
M. Sewi{\l}o$^{13,14}$,
J. Th. van Loon$^{15}$
}\vspace{0.5cm}\\
\parbox{\textwidth}{$^{1}$ Institute of Astronomy, University of Cambridge, Madingley Road, Cambridge CB3 0HA, UK\\
$^{2}$ Laboratoire AIM, CEA, Universit\'{e} Paris Diderot, IRFU/Service d'Astrophysique, Bat. 709, 91191 Gif-sur-Yvette, France\\
$^{3}$ Argelander-Institut f\"ur Astronomie, Auf dem H\"ugel 71, D-53121 Bonn, Germany\\
$^{4}$ Observatoire astronomique de Strasbourg, Universit\'e de Strasbourg, CNRS, UMR 7550, 11 rue de l'Universit\'e, F-67000 Strasbourg, France\\
$^{5}$ Steward Observatory, University of Arizona, Tucson, AZ 85721, USA\\
$^{6}$ Raytheon Company, 1151 East Hermans Road, Tucson, AZ 85756, USA\\
$^{7}$ Department of Astrophysics, Nagoya University, Chikusa-ku, Nagoya 464-8602 , Japan\\
$^{8}$ Sterrewacht Leiden, Leiden University, P.O. Box 9513, NL-2300 RA Leiden, The Netherlands\\
$^{9}$ National Astronomical Observatory of Japan, 2-21-1, Osawa, Mitaka, Tokyo, 181-8588, Japan\\
$^{10}$ 314 Physics Building, Department of Physics and Astronomy, University of Missouri, Columbia, MO 65211, USA\\
$^{11}$ Space Telescope Science Institute, 3700 San Martin Drive, Baltimore, MD 21218\\
$^{12}$ Departamento de Astronomia, Universidad de Chile, Casilla 36-D, Santiago, Chile\\
$^{13}$ Johns Hopkins University, Department of Physics and Astronomy, Homewood Campus, Baltimore, MD 21218, USA\\
$^{14}$ Space Science Institute,  4750 Walnut St. , Suite 205, Boulder, CO 80301, USA\\
$^{15}$ School of Physical \& Geographical Sciences, Lennard-Jones Laboratories, Keele University, Staffordshire ST5 5BG, UK
}
}

\maketitle{}

%===============================================================
% Abstract
%===============================================================
 
\begin{abstract}
Low-metallicity galaxies exhibit different properties of the interstellar medium (ISM) compared to nearby spiral galaxies. Obtaining a resolved inventory of the various gas and dust components of massive star forming regions and diffuse ISM is necessary to understand how those differences are driven. We present a study of the infrared/submillimeter (submm) emission of the massive star forming complex N158-N159-N160 located in the Large Magellanic Cloud. Combining observations from the \spitz\ Space Telescope (3.6-70 \mic), the \hersc\ Space Observatory (100-500 \mic) and LABOCA (on APEX, 870 \mic) allows us to work at the best angular resolution available now for an extragalactic source (a few parsecs for the LMC).
We observe a remarkably good correlation between the \hersc\ SPIRE and LABOCA emission and resolve the low surface brightnesses emission. We use the \spitz\ and \hersc\ data to perform a resolved Spectral Energy Distribution (SED) modelling of the complex. Using modified blackbodies, we derive an average ``effective" emissivity index of the cold dust component $\beta$$_c$ of 1.47 across the complex. If $\beta$$_c$ is fixed to 1.5, we find an average temperature of $\sim$27K (maximum of $\sim$32K in N160). We also apply the \citet{Galliano2011} SED modelling technique (using amorphous carbon to model carbon dust) to derive maps of the star formation rate, the grain temperature, the mean starlight intensity, the fraction of Polycyclic Aromatic Hydrocarbons (PAH) or the dust mass surface density of the region. We observe that the PAH fraction strongly decreases in the H\,{\sc ii} regions we study. This decrease coincides with peaks in the mean radiation field intensity map. The dust surface densities follow the far-infrared distribution, with a total dust mass of 2.1 $\times$ 10$^4$ \msun\ (2.8 times less than if carbon dust was modelled by standard graphite grains) in the resolved elements we model. We also find a non-negligible amount of dust in the region called ``N159 South", a molecular cloud that does not show massive star formation. We also investigate the drivers of the \hersc/PACS and SPIRE submm colours and find that the submm ratios correlate strongly with the radiation field intensity and with the near and mid-IR surface brightnesses equally well. Comparing our dust map to H\,{\sc i} and CO observations in N159, we then investigate variations in the gas-to-dust mass ratio (G/D) and the CO-to-H$_2$ conversion factor X$_{CO}$. A mean value of G/D$\sim$356 is derived when using X$_{CO}$ = 7$\times$10$^{20}$ H$_2$ cm$^{-2}$ (K km s$^{-1}$)$^{-1}$ \citep{Fukui2009}. If a constant G/D across N159 is assumed, we derive a X$_{CO}$ conversion factor of 5.4$\times$10$^{20}$ H$_2$ cm$^{-2}$ (K km s$^{-1}$)$^{-1}$. We finally model individual regions to analyse variations in the SED shape across the complex and the 870 \mic\ emission in more details. No measurable submm excess emission at 870 \mic\ seems to be detected in these regions.
\end{abstract}
  
\begin{keywords}
galaxies: ISM --
		galaxies:dwarf--
		galaxies:SED model --
		ISM: dust --
		submillimeter: galaxies
\end{keywords}

%     \authorrunning{Galametz et al}
%     \titlerunning{Studying the dust in the N159 region with SAGE, HERITAGE and LABOCA}

%===============================================================
%Introduction
%===============================================================

\section{Introduction}

As potential templates of primordial environments, low-metallicity galaxies are keystones to understand how galaxies evolve through cosmic time. Studying the Interstellar Medium (ISM) of low-metallicity galaxies is a necessary step to get a handle on the interplay between star formation and the ISM under conditions characteristic of the early universe. Low-metallicity galaxies also have quite different infrared (IR) Spectral Energy Distributions (SEDs) than solar or metal-rich environments. For instance, their aromatic features are diminished compared to dustier galaxies \citep{Madden2006,Engelbracht2008}. The paucity of aromatic features is usually attributed to the hardness of the radiation field in low-metallicity environments, destructive processes such as supernova-driven shocks effects \citep{Galliano2003,Galliano2005} or delayed injection of carbon dust by asymptotic giant branch (AGB) stars \citep{Dwek1998,Galliano_Dwek_Chanial_2008} in dwarf galaxies. More relevant to the present study, the SEDs of low-metallicity galaxies often exhibit a flattening of their submillimeter (submm) slope or a submm excess \citep{Bottner2003, Galliano2003,Galliano2005,Marleau2006, Bendo2006,Galametz2009, Galametz2011}, namely a higher emission beyond 500 \mic\ than that extrapolated from IR observations and standard dust properties (Milky Way dust for instance). The origin of this excess is still highly debated. These results highlight the importance of a complete coverage of the thermal dust emission of low-metallicity objects to get a handle on the overall dust population distribution and properties in these environments. Combining gas and dust tracers will also allow us to understand how the matter cycles and the star formation processes evolve with galaxy properties.

The Large Magellanic Cloud (LMC) is our nearest low-metallicity neighbour galaxy \citep[$\sim$ 50kpc;][]{Feast1999}, well studied at all wavelengths. This proximity enables us to study in detail the physical processes at work in the ISM of the galaxy and to individually resolve its bright star-forming regions. These structures can unbiasedly be isolated due to the almost face-on orientation \citep[23-37$^{o}$][]{Subramanian2012}. Furthermore, the low interstellar extinction along the line of sight facilitates the interpretation of the physical conditions of these star-forming regions compared to Galactic studies strongly affected by this extinction. Thus, the irregular morphology and low-metallicity of the LMC \citep[$\sim$ 1/2 \zsun;][] {Pagel2003} make it a perfect laboratory to study the evolution in the physical properties of the ISM of galaxies and the influence of metal enrichment on their star-forming activity.

Before the \spitz\ {\it Space Telescope} (\spitz) observations, the infrared studies on the LMC suffered from a lack of wavelength coverage or spatial resolution to quantify crucial physical parameters of the ISM properties such as the equilibrium temperature of the big grains, the spatial distribution of the dust grain populations or the interstellar radiation field. A study of the extended infrared emission in the ISM of the LMC was performed by \citet{Bernard2008} as part of the \spitz\ Legacy Program SAGE \citep[Surveying the Agents of a Galaxy's Evolution,][]{Meixner2006} project using \spitz\ observations. They found disparities between the overall shape of the SED of the LMC and that of the Milky Way (MW), namely a different mid-infrared (MIR) shape. They also found departures from the linear correlation between the FIR optical depth and the gas column density. Using \spitz\ FIR spectra ($\lambda$ = 52-93\mic), the studies of \citet{VanLoon2010b} also allowed comparisons of compact sources in the LMC and its neighbor, the Small Magellanic Cloud (SMC), that has an even lower metallicity. Their results indicate that while the dust mass differs in proportion to metallicity, the oxygen mass seems to differ less. The photo-electric effect is indistinguishably efficient to heat the gas in both clouds. The SMC finally presents evidence of reduced shielding and reduced cooling.

At submm wavelengths, the LMC was observed by \citet{Aguirre2003} with the TopHat instrument, a balloon-borne telescope \citep{Silverberg2003} from 470 \mic\ to 1.2 mm. They constrain the FIR regime with DIRBE (Diffuse Infrared Background Experiment) observations at 100, 140 and 240 \mic\ and estimated an average dust temperature for the LMC of T=25.0$\pm$1.8K. Using DIRBE and ISSA (IRAS Sky Survey Atlas) observations, \citet{Sakon2006} found that the submm emission power law index (often referred to as $\beta$) is smaller in the LMC than in the MW and that the 140 and 240 \mic\ fluxes seem to deviate from the model predictions, in particular on the periphery of supergiant shells. This excess was modelled by a very cold dust component with temperatures $<$ 9K, even if their lack of submm constraints prevented them to unbiasedly conclude that cold dust was the explanation for the excess. Using revised DIRBE, WMAP and COBE maps, \citet{israel2010} construct the global SED of the LMC to confirm a pronounced excess emission at millimeter and submm wavelengths, i.e. more emission than expected from a submm slope with $\beta$=2. Different hypotheses for this global excess are tested in \citet{Bot2010_2}, \citet{Bot2010}, \citet{Galliano2011} (both rule out for instance the ``very cold dust" hypothesis) or \citet{Planck_collabo_2011_MagellanicClouds} (that test the ``spinning dust" hypothesis).

We more specifically focus the present study on the N158-N159-N160 complex, a group of H\,{\sc ii} regions ($\sim$ 400pc from North to South) located in the LMC, $\sim$ 500pc south of the 30 Doradus (30Dor) massive star-forming region \citep[catalogued by][]{Henize1956}. The complex is intensively observed in IR, H\,{\sc i} or CO. We have obtained observations of the complex with the LABOCA instrument at 870 \mic, probing the coldest phases of dust at a resolution of 19\arcsec. We note that this analysis presents the first study of LABOCA observation in the LMC. We refer to \citet{Bot2010_2} for a detailed analysis of LABOCA observations of giant molecular clouds in the south-west region of the SMC. In addition to LABOCA data, the {\it Herschel Space Observatory} (\hersc) now allows us to probe the thermal dust emission with a coverage of the SED from 70 \mic\ up to 500 \mic, thus to sample the IR peak and the submm slope of nearby galaxies. The whole LMC has been mapped with \hersc\ as part of the HERITAGE project (HERschel Inventory of The Agents of Galaxy Evolution; \citealt{Meixner2010}, Meixner et al., submitted to AJ). The good resolution of the instruments on-board \hersc\ favours the exploration of the local properties of the ISM with a resolution of \hersc/SPIRE at 500 \mic\ similar to that of \spitz/MIPS 160 \mic\ (FWHM of the PSF=36\arcsec), leading, for the LMC, to ISM resolution elements of $\sim$9 pc at SPIRE 500 \mic.

%------------------------------------------------------------------------------------------------------------------------------------------
\begin{figure*}
    \centering
    \begin{tabular}{ m{8cm}m{9cm} }
    \hspace{-20pt}
     \includegraphics[width=9.5cm]{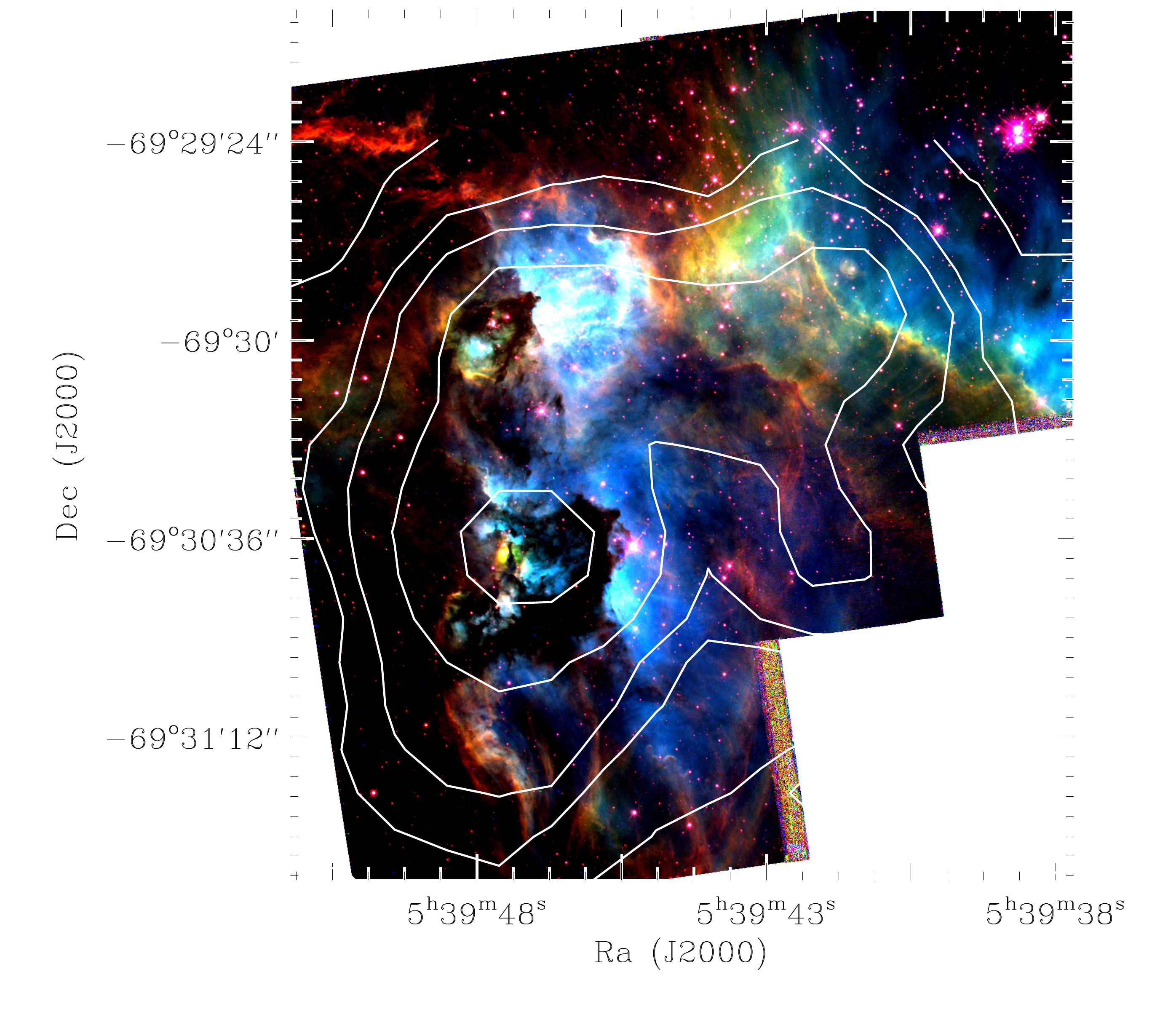} &
       \includegraphics[width=9.5cm]{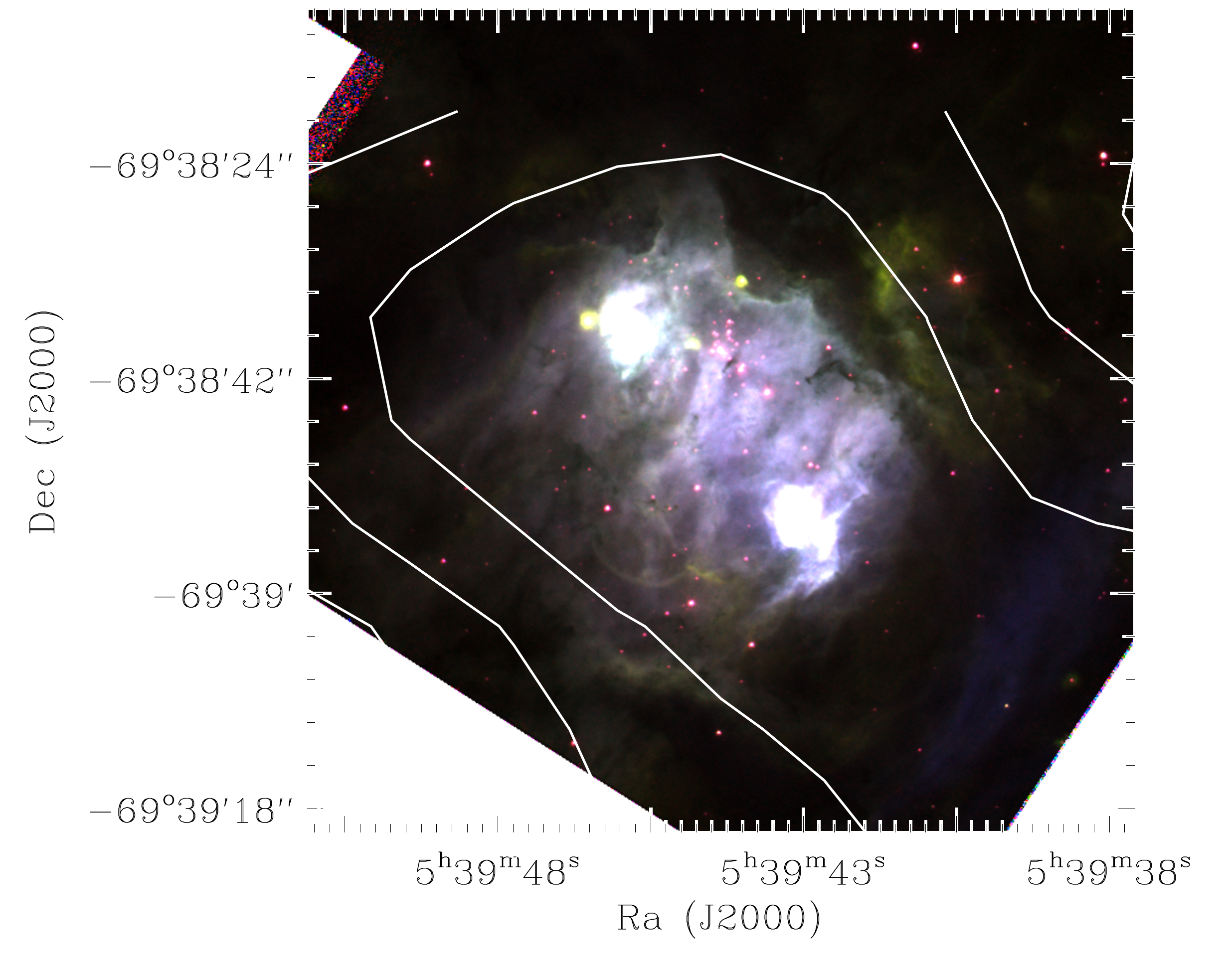} \\
         \end{tabular}
    \caption{The N158-N160-N159 complex is located in the LMC ($\sim$ 500pc south of 30Dor). We show here three-color compositions of the two regions N158 and N160 produced with images taken with HST. For N158 ({\it left}), we use a combination of the narrow-band filters F502N/[OIII] (blue), F656N/H$\alpha$ (green) and F673N/[SII] (red). For N160 ({\it right}), we use a composition of the narrow-band filters F502N/[OIII] (blue), F656N/H$\alpha$ (green) and F487N/H$\beta$ (red). LABOCA 870 \mic\ contours are overlaid on both images. Contours are given for 0.018, 0.06, 0.09, 0.15 and 0.3 Jy~beam$^{-1}$ on the left image, 0.15 and 0.3 Jy~beam$^{-1}$ on the right image. North is up, east is left.}
    \label{HST_LABOCA}
\end{figure*}
%------------------------------------------------------------------------------------------------------------------------------------------

This paper thus presents a study of the IR to submm thermal emission of star-forming regions in various evolutionary states at the best resolution available now. This allows us to investigate how properties and physical conditions observed on large scales are physically linked to specific environments on a more resolved scale.  In Section 2, we describe the N158-N159-N160 complex and the observations and data reduction of the \lab, \spitz\ and \hersc\ data. We present the SED modelling techniques we apply to the \spitz\ and \hersc\ data on resolved scales in Section 3. In Section 4, we analyse the results of our modelling, namely the distribution of the dust temperatures, the mean stellar radiation field intensities, the fraction of Polycyclic aromatic hydrocarbons (PAH) or the star formation rates. We also study the dependence of submm colours on various parameters and compare our dust mass map with the available gas tracers for the N159 region. A SED modelling of individual regions across the complex including the 870 \mic\ data in the fitting procedure is described in Section 5. We summarize the main conclusions of the analysis in Section 6.

%------------------------------------------------------------------------------------------------------------------------------------------
\begin{figure*}s
    \centering
    	\vspace{20pt}
       \includegraphics[width=18.5cm, height=21cm]{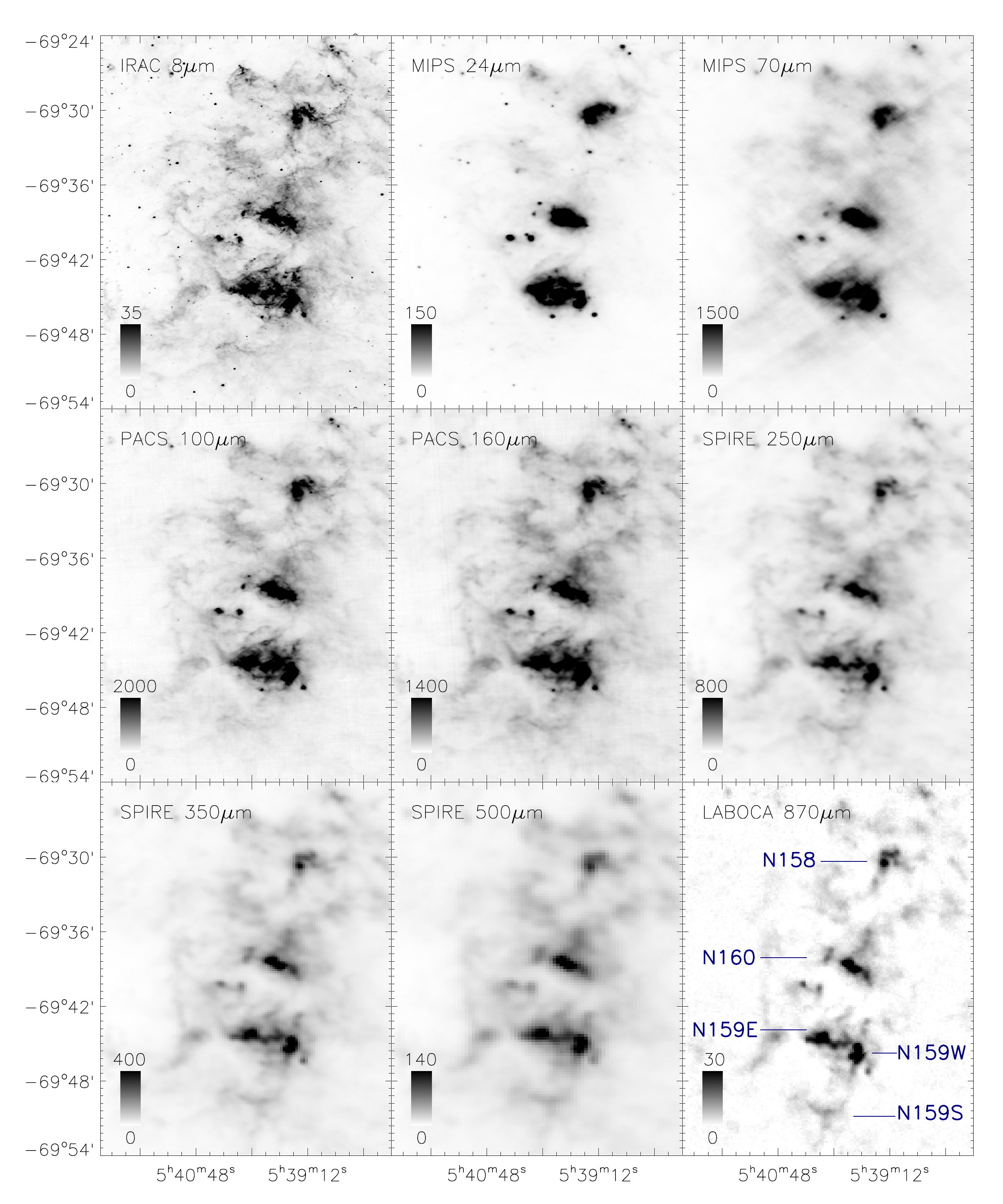}     \\    
     \caption{Images of the N158-N159-N160 star-forming complex observed from 8 to 870 \mic. North is up, east is left and coordinates (RA, Dec) are given in equinox J2000. We provide the intensity scale in MJy~sr$^{-1}$ for each panel.}
    \label{Spitzer_Herschel_LABOCA_maps}
\end{figure*}
%------------------------------------------------------------------------------------------------------------------------------------------

%===============================================================
\section{A multi-wavelength dataset}

\subsection{The complex}

The N158-N159-N160 complex ($\sim$400pc from North to South) is located in the LMC at $\sim$500pc south of the 30Dor complex, the most massive star-forming region of the Local Group \citep{Kennicutt_Hodge_1986}.  Figure~\ref{HST_LABOCA} shows three-color compositions of the two regions N158 and N160 produced with images taken by the Wide Field Planetary Camera 2 (WFPC2) on board the Hubble Space Telescope (HST). Data are retrieved from the Hubble Legacy Archive (http://hla.stsci.edu/, Program IDs: 11807 and 8247 for N158 and N160 respectively). We overlay LABOCA 870 \mic\ contours on both images for comparison between the spatial resolutions. The HST/WFPC2 observations of N158, N159 and N160 are discussed in details in \citet{Fleener2010}, \citet{Heydari1999} and \citet{Heydari2002} respectively. \\

\vspace{-7pt}
{\it N158 - }This H\,{\sc ii} region (Fig.~\ref{HST_LABOCA}, left) possesses an elongated structure in the N-S direction. Two OB associations were first discovered in the region by \citet {Lucke1970} who catalogued stellar associations across the whole LMC. One is associated with the southern part of N158, one with the north superbubble. Several studies have shown that this superbubble is dominated by a young stellar population of 2 to 6 Myr \citep{Testor1998,Dunne2001}.\\

\vspace{-7pt}
{\it N159 - }Observations of N159 in the K band by \citet{Gatley1981} led to the discovery of the first extragalactic protostar. The stellar content of sub-structures of N159 was analysed in more detail by \cite{Deharveng1992} at optical wavelengths and \citet{Meynadier2004} using JHK photometry. N159 presents characteristics of on-going star formation activity such as protostars, masers, young massive stars as well as High-Excitation Blobs (HEB), namely nebulae with turbulent media and strong stellar winds interacting with the ambient ionized gas. The Kuiper Airborne Observatory (KAO) has been used to probe the complex at far-infrared (FIR) wavelengths for the first time \citep{Werner1978,Jones1986}. Later, \citet{Jones2005} used \spitz/IRAC observations to follow the protostars in N159 and constrain the nature of the upper end of the initial mass function (IMF) in the LMC. \\

\vspace{-7pt}
{\it N160 - }As in N159, massive star formation is well evolved in N160 (Fig.~\ref{HST_LABOCA}, right), with parent clouds that have mostly been dissipated. The region is associated with H\,{\sc ii} regions, young stellar clusters and water masers \citep{Lazendic2002,Oliveira2006}. Using JHK photometry, \citet{Nakajima2005} showed that of the 11 optical stellar clusters and associations detected in the N159/N160 complex (ages listed by \citealt{Bica1996}), one has an age superior to 10 Myr and belongs to N160, the rest are younger than 10 Myr, with 3 clusters showing populations younger than 3 Myr. We refer to \citet{Carlson2012} for a study of the physical properties and evolutionary stages of YSOs in N160 and to \citet{VanLoon2010a} for a detailed study of FIR fine-structure cooling lines of compact sources in the 3 H\,{\sc ii} regions.\\

The N158-N159-N160 complex belongs to the largest molecular cloud complex of the LMC called ``the molecular ridge" and accounts for $\sim$30$\%$ of the total molecular mass of the galaxy \citep{Mizuno2001}. Several analyses have suggested a dissipated sequential cluster formation within N158, N160 and N159 as well as a large-scale sequential cluster formation over the entire complex \citep{Bolatto2000, Nakajima2005} and more globally from the 30Dor complex to the southern region of N159 where the large southward CO ridge of 30Dor begins \citep[][]{Cohen1988,Fukui1999}. Near-infrared (NIR) to FIR observations with {\it 2MASS} and \spitz\ highlighted the paucity of star-formation across the ridge, with a star-formation activity distributed in lower luminosity regions than in the 30Dor complex \citep{Indebetouw2008}. Several studies have suggested alternative scenarios to the standard self-propagating star formation to explain this trend, for instance that the sequential star formation could be induced in bow-shocks formed at the leading edge of the LMC \citep{deBoer1998}. 

The N159 H\,{\sc ii} region harbours two giant molecular clouds (GMCs), known as ``N159 East" (N159E) and ``N159 West" (N159W) labeled in Fig.~\ref{Spitzer_Herschel_LABOCA_maps}. \citet{Chen2009} suggest that the current star formation in the N159E is probably triggered by H\,{\sc ii} regions in expansion in the molecular cloud, while the massive stars of N159W are forming spontaneously. Another GMC was observed in the south of N159 (N159S). Performing a comparative study between [CII], CO, and FIR observations, \citet{Israel1996_2} find that the clouds N160 and N159S are two opposite extremes with respect to N159 (East and West), with molecular gas in N160 more photo-dissociated than in N159. N159S shows a much lower level of star formation than N159W and N159E \citep{Bolatto2000,Rantakyro2005}, with only a few very faint diffuse H\,{\sc ii} regions and no OB associations \citep{Chen2010}. The detection of candidates for Herbig Ae/Be stars suggest that the cluster formation could have just begun in N159S \citep{Nakajima2005}. The atomic and molecular gas distributions across the complex are discussed in more details in Section 4.5.

\subsection{LABOCA observations and data reduction}

The full region of N158, N159 and N160 was mapped with the \lab\ instrument operating at 870 \mic\ and installed on the Cassegrain cabin of the Atacama Pathfinder EXperiment (APEX) telescope in the Chajnantor site in Chile's Atacama desert. \lab\ is a bolometer array of 295 channels  with a field of view of the full array of 11\farcm4 . 
The full width half maximum (FWHM) of its point spread function (PSF) is 19\farcs2 $\pm$ 0.3. Detectors of LABOCA are positioned with a double beam channel separation (36\farcs) and the field of view is thus under-sampled. The mapping was carried out with a raster of spiral pattern to obtain an homogenous fully sampled map. A total of 35.5 hrs of observations were obtained in August 2008 (Program ID: O-081.F-9329A-2008 - Principal Investigator: Sacha Hony). Calibration was performed through the observations of Mars and secondary calibrators: PMNJ0403-8100, PKS0537-441, B13134, V883-ORI, N2071IR and VY-CMa. The atmospheric attenuation was determined via skydips every hour\footnote{See http://www.apex-telescope.org/bolometer/laboca/calibration/ for details on the calibration and archive tables of sky opacities and calibration factors.}. 

The data reduction was performed with the BoA reduction package (BOlometer Array Analysis Software)\footnote{BoA was developed at MPIfR (Max-Planck-Institut fŸr Radio Astronomy, Bonn, Germany), AIfA (Argelander-Institut f{\"u}r Astronomie, Bonn, Germany), AIRUB (Astronomisches Institut der Ruhr-Universit{\"a}t, Bochum, Germany), and IAS (Institut d'Astrophysique Spatiale, Orsay, France)}. The main steps of the reduction of the time-ordered data stream of each channel and scan are: flat fielding, calibration, opacity correction, dead or noisy channels removal, removal of the correlated noise on the global array as well as correlated noise induced by the coupled electronics of the detectors (amplifier boxes or cables), flagging of stationary points or data taken outside reasonable telescope scanning velocity and acceleration limits, $^3$He temperature drift correction, median baseline removal and despiking. Individual reduced scans are then co-added.

The presence of very bright structures often leads to negative artefacts on their surrounding. These structures are ``created" numerically during the correlation noise or median baseline steps. It is possible to reduce these artefacts using an iterative process during the data reduction. Once data are calibrated, we use our reduced image to create a ``source model" by isolating the pixels superior to a given signal-to-noise (5 for the first iterations and decreasing progressively to 2). The model grows automatically to avoid isolated pixels in the masks. We subtract the source model from the data and rerun the reduction pipeline. We add the model at the end of the new reduction to obtain a new map. This map is used to build a new model, input for the following iteration. The process is repeated until the reduction converges. Those iterations lead to a significant retrieval of faint extended emission around the bright structures. Weight maps are derived during the final mapping step from which we derive rms and signal-to-noise maps. The final 45\arcmin $\times$ 35\arcmin~\lab\ image has a pixel size of 9\farcs1, with a final rms of 7.8 mJy~beam$^{-1}$.

%The major sources of uncertainty are the calibration uncertainty. We estimate the average uncertainty levels to be of $\sim$ 15$\%$. 

\subsection{Herschel Data}

We obtained \hersc\ data for the N158-N159-N160 complex from the HERITAGE project, a programme dedicated to the observations of the two Magellanic Clouds and a part of the Magellanic Bridge. We refer to \citet{Meixner2010} and the HERITAGE overview of Meixner et al. (submitted) for a detailed description of the observing strategy of the HERITAGE project and the data reduction of the data. We provide a short summary here. The LMC was mapped in two bands with PACS \citep[Photodetector Array Camera and Spectrometer;][]{Poglitsch2010} at 100 and 160 \mic, with respective FWHM of the Point-Spread Functions (PSF) of $\sim$7\farcs7 and $\sim$12\arcsec. Data are processed in HIPE 7.0 (\hersc\ Interactive Processing Environment) from Level 0 to Level 1 following the standard pipeline described in the PACS data reduction guide. Several steps of baseline subtraction (to take 1/$\sqrt[]{f}$ noise and drifts with time into account) and deglitching (to correct for jumps caused by cosmic ray hits) are then applied to the data. PACS 100 and 160 \mic\ timelines are finally mapped using the PhotProject HIPE procedure.
The LMC was also observed with \hersc/SPIRE \citep[Spectral and Photometric Imaging Receiver;][]{Griffin2010} at 250, 350 and 500 \mic, with respective FWHMs of their PSFs of 18\arcsec, 25\arcsec and 36\arcsec. Data were reduced in HIPE 7.0, including additional routines to subtract the background, adjust the astrometry, apply deglitching steps, remove residuals from temperature drifts or mask discrepant data. 
We expect residual foreground emission as pointed out by \citet{Bernard2008}. Using the HI data cube of the LMC restricted to velocity ranges matching those of the Galaxy, \citet{Galliano2011} have quantified this contamination to contribute to $\sim$1 $\%$ of the IR power, i.e. smaller than \hersc\ flux uncertainties. We thus consider the foreground contamination to be negligible in this study.

\subsection{Spitzer Data}

The \spitz\ observations were performed as part of the SAGE project (Surveying the Agents of a Galaxy's Evolution) using both instruments IRAC \citep[InfraRed Array Camera;][]{Fazio2004} and MIPS \citep[Multiband Imaging Photometer;][]{Rieke2004}. The IRAC bands cover 3.6, 4.5, 5.8, and 8 \mic\ with FWHMs of the PSFs $<$2\arcsec. The MIPS bands cover 24, 70 and 160 \mic\ with a FWHM of the PSFs of 6\arcsec, 18\arcsec\ and 40\arcsec\ respectively. A description of the observing strategy and data reduction can be found in \citet{Meixner2006}. We also refer to \citet{Jones2005} for a very detailed overview of the IRAC observations of N159 and its various components.
Because of their lower resolution (40\arcsec) compared to the \hersc/PACS 160 \mic\ maps (12\arcsec), we do not use the MIPS 160 \mic\ maps in this study. Meixner et al. (submitted) indicate a good agreement of the MIPS and PACS calibrations in the linear range of MIPS.

%------------------------------------------------------------------------------------------------------------------------------------------
\begin{figure}
    \centering
    \begin{tabular}{c }
    \hspace{-20pt}
     \includegraphics[width=8.3cm]{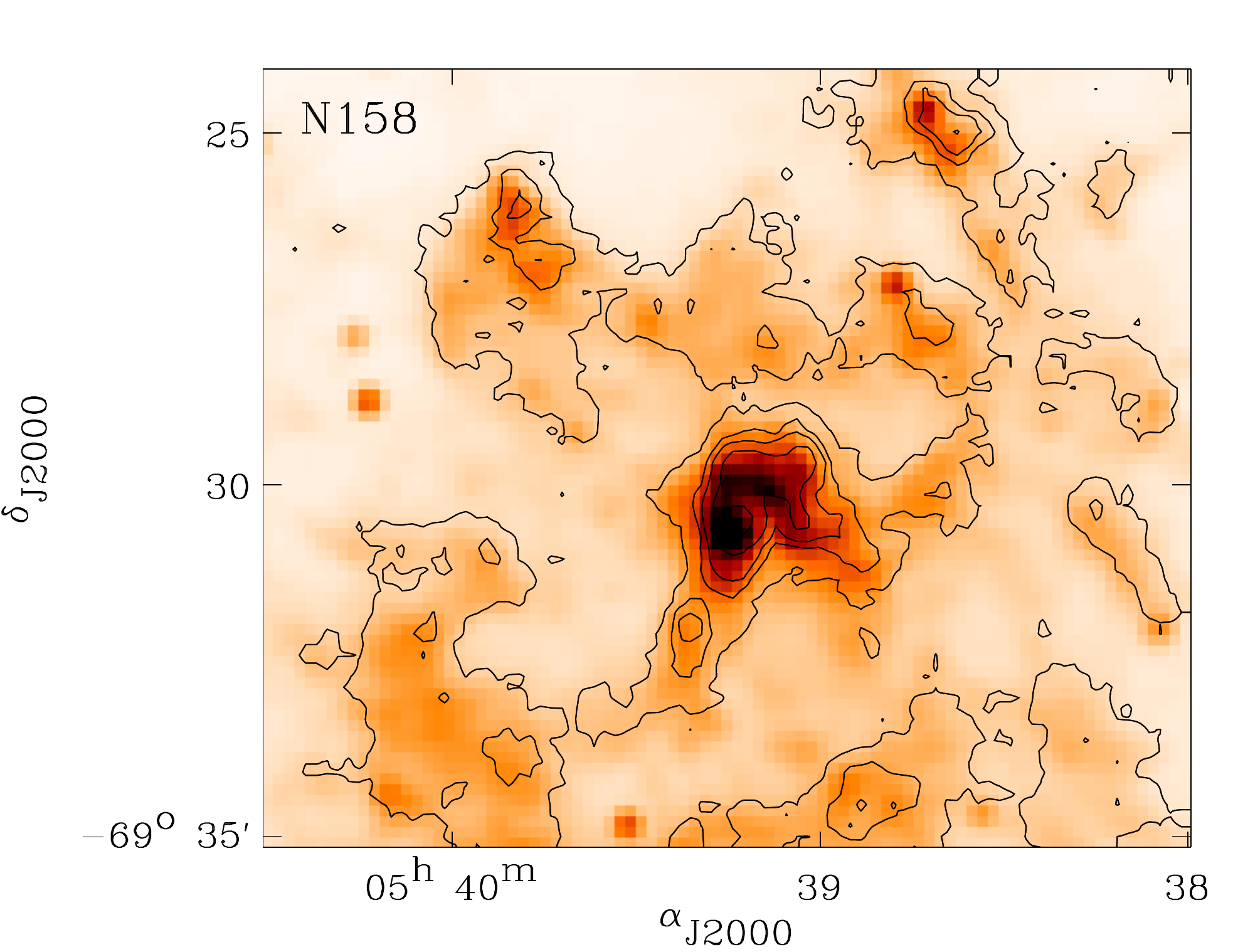}   \vspace{-12pt}\\
     \hspace{-23pt}
      \includegraphics[width=8.3cm]{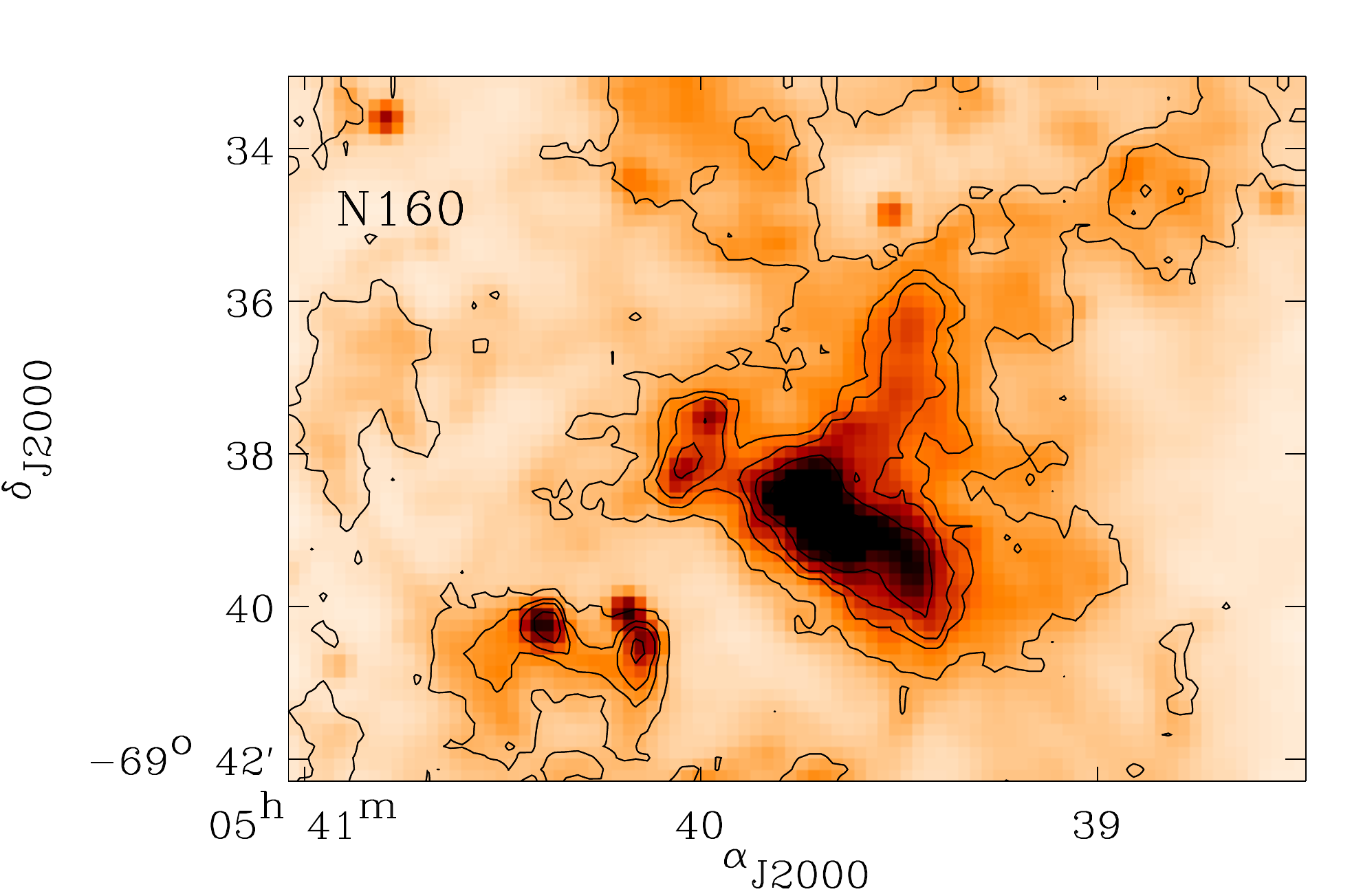}    \vspace{-12pt}\\
     \hspace{-23pt}
      \includegraphics[width=8.8cm]{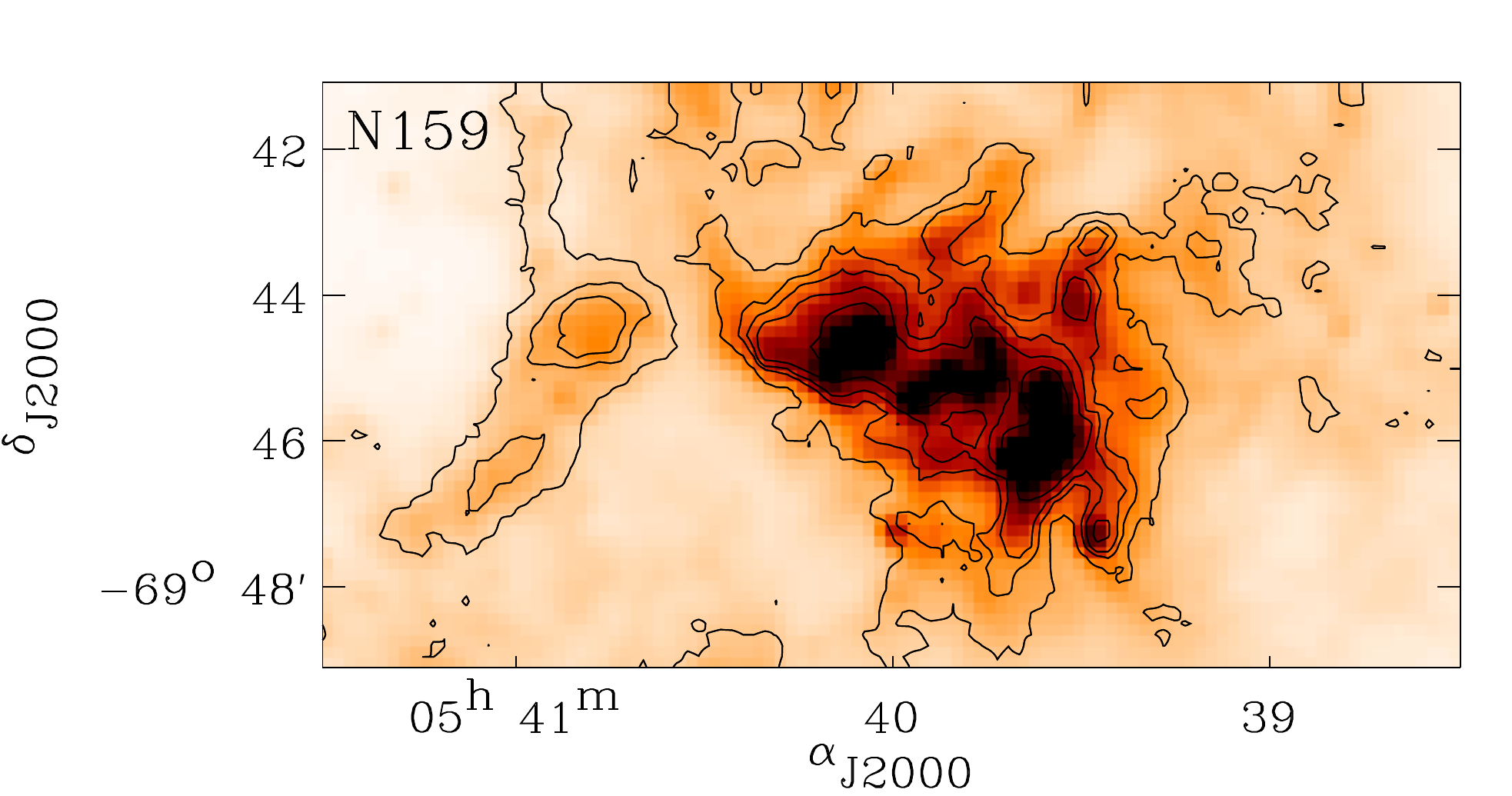}     \vspace{-13pt}\\
     \hspace{-23pt}
      \includegraphics[width=6.8cm]{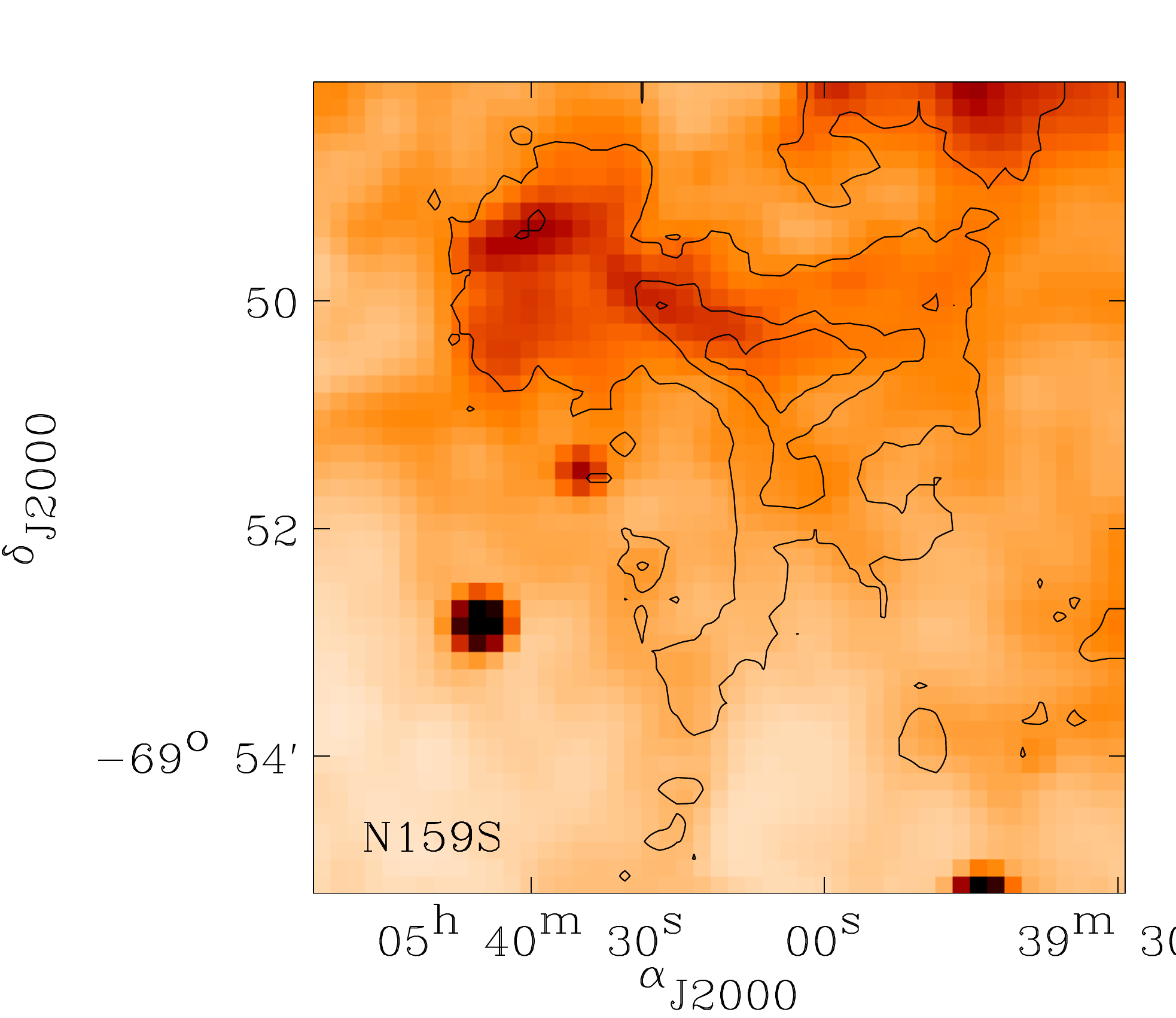}  \vspace{-12pt}\\
         \end{tabular}
         \vspace{10pt}
    \caption{Details on the IR/submm emission in the regions. LABOCA 870 \mic\ contours are overlaid on the IRAC 8 \mic\ observation with, from top row to bottom row N158, N160, N159 and N159S. The IRAC 8 \mic\ map is convolved to the LABOCA resolution. Contours are given for 0.018, 0.06, 0.09, 0.15 and 0.3 Jy/beam. North is up, east is left.}
    \label{IRAC_LABOCA}
\end{figure}
%------------------------------------------------------------------------------------------------------------------------------------------

\subsection{Variation of the thermal dust emission with wavelength}

We present a multi-wavelength composition of the N158-N159-N160 region in Fig.~\ref{Spitzer_Herschel_LABOCA_maps}, with, from top to bottom IRAC 8 \mic, MIPS 24 and 70 \mic, PACS 100 and 160 \mic, the three SPIRE bands (250, 350 and 500 \mic) and the \lab\ map at 870 \mic. This figure shows how the thermal dust emission is evolving with wavelength. The 8 \mic\ observation mostly traces the aromatic feature emission from PAHs. The MIPS 24 \mic\ emission is more compact than that of the other bands. Tracing the hot dust, the 24 \mic\ emission is indeed confined to the compact star-forming sites. The cool dust (10-40K) is now traced by the \hersc\ PACS and SPIRE observations. The \lab\ observations are finally probing for the first time the coldest phases of dust at a resolution of 19\arcsec\ (4.6pc for the LMC). 

We clearly resolve the emission of the bright star-forming regions of the complex with LABOCA at 870 \mic, but also detect and resolve the diffuse emission at low surface brightnesses. The distribution of the 870 \mic\ emission is spatially coherent with that of the SPIRE maps, both in the H\,{\sc ii} regions and in the diffuse medium. To allow spatial comparison between the 870 \mic\ maps and the IRAC 8 \mic\ map, predominantly tracing PAHs, we convolve the IRAC 8 \mic\ map to the LABOCA resolution using the low-resolution kernel created by G. Aniano from the PSFs of IRAC and \lab\footnote{http://www.astro.princeton.edu/$\sim$ganiano/Kernels.html for kernels and details on their construction.} and observe a spatial coincidence between the 8\mic\ map and the 870\mic\ contours (Fig.~\ref{IRAC_LABOCA}) already observed by \citet{Haas2002} for a wide range of galaxy types. On one hand, this correlation suggests that the PAH excitation could be caused by a widespread distribution of stellar FUV-emitters and not only linked with star formation \citep{Whaley2009}. On the other hand, we expect the 870 \mic\ emission, namely the cold dust emission, to predominantly originate from largely shielded interiors of molecular clouds. Their surfaces are known to produce the PAH features \citep{Sauvage2005}, which could explain their association with the 870\mic\ emission.

We finally note the peculiar behaviour of the southern region of N159 (N159S in Fig.~\ref{Spitzer_Herschel_LABOCA_maps}). As observed in Fig.~\ref{Spitzer_Herschel_LABOCA_maps} and Fig.~\ref{IRAC_LABOCA}, the peak of emission moves from 8 to 870 \mic, with an offset of  2.7\arcmin\ toward the west for the 870 \mic\ emission. We will see further in this paper that the peak of the 870 \mic\ emission (which also dominates the emission at SPIRE wavelengths) spatially coincides with the molecular gas reservoir usually referred to as ``N159 South" (N159S). This region was indeed observed in CO \citep{Bolatto2001,Mizuno2010} and is the brightest [C{\sc i}] source of N159 \citep{Bolatto2000}. We provide details on previous observations of gas tracers (H\,{\sc i}, CO etc.) in the complex in Section 4.5.

\subsection{Convolution and background subtraction}

We convolve the IRAC, MIPS, PACS and SPIRE 250 and 350 \mic\ maps to the lowest resolution available, namely that of SPIRE 500 \mic\ (FWHM: 36\arcsec). We used the convolution kernels developed by \citet{Aniano2011}, which are rotationally symmetric and optimized to avoid high-frequency numerical noise (induced by the filtering steps when one works in the frequency domain). The LABOCA 870 \mic\ map is convolved to the SPIRE 500 \mic\ resolution using the low resolution kernel created by G. Aniano from the PSFs of both SPIRE 500 \mic\ and \lab\ (see Section 2.5). 
For each image, we estimate the background by masking the emission linked with the complex and fitting the distribution of the remaining pixels by a Gaussian. The peak value of this Gaussian is used as a background value per pixel and subtracted from the maps. In the following study, we restrict ourselves to resolved elements with a signal-to-noise superior to 1-$\sigma$ in the SPIRE bands.

%===============================================================

 \section{A resolved SED modelling}

We use data from 3.6 to 500 \mic\ to obtain maps of the dust properties within the star-forming complex. Because the data reduction of the LABOCA data might still lead to removal of some of the faintest diffuse emission, we do not use the 870 \mic\ map as a constraint for the resolved modelling. We derive average cold temperatures on a resolved basis using a two-temperature (two modified blackbodies) fitting procedure. We then apply a more realistic model \citep{Galliano2011} in order to investigate the distribution of the star formation rate, radiation field properties or dust properties (mass, PAH fraction) across the star-forming complex. We describe the two resolved SED fitting procedures in this section and analyse the maps of the physical parameters we obtained in Section 4. In Section 5, we perform a SED modelling of 24 individual regions across the complex (same model) to derive integrated properties on those regions and investigate the 870 \mic\ thermal emission separately.

\subsection{Modified blackbody fitting}

If we assume that dust grains are in thermal equilibrium with the radiation field, we can obtain a simple estimate of the cold dust temperature across the complex by fitting a modified blackbody (MBB) to the dust thermal emission. It is important to keep in mind that in reality, each resolution element contains dust grains with varied properties along the line of sight. The temperatures derived using this method should thus be considered as luminosity-weighted average values. We choose to fit our 24 to 500\mic\ data with a two-temperature (warm+cold) model. This two-MBB fitting is a minimum requirement to account for the contribution of warm dust to the MIR emission (and estimate the contribution of warm dust and single-photon-heating) to the 70 \mic\ emission) and avoid an overestimation of the cold dust temperatures. 

We fit our data with a model of the form:
\begin{eqnarray}
L_{\nu}(\lambda, T_w, T_c, \beta_c)  = A_w~\lambda^{-2} B_{\nu}(\lambda,T_w) + A_c~\lambda^{- \beta_c} B_{\nu}(\lambda,T_c)
    \label{eq1}
\end{eqnarray}

\noindent with {\it T$_{w}$} and {\it T$_{c}$} the temperature of the warm and cold component, {\it $\beta$$_{c}$} the emissivity index of the cold component and {\it B$_{\nu}$} the Planck function. The A$_w$ and A$_c$ are scaling coefficients that account for the dust masses of each component. The emissivity index of the warm dust component is fixed to 2, a standard approximation of the opacity in the \citet{Li_Draine_2001} dust models. We convolve the model with the instrumental spectral responses of the different cameras to derive the expected photometry. The fit is performed using the IDL function MPCURVEFIT \citep[][Levenberg-Marquardt least-squares fit]{Markwardt2009}. We take the uncertainties on flux measurements into account to weight the data during the fitting procedure (normal 1/error$^{2}$ weighting). \\

{\it Choice of emissivity - }In order to derive an estimate of the emissivity index value of the complex, we first let the emissivity index of the cold dust component vary ($\beta$$_c$ free) and perform the two-MBB fitting for each ISM resolved elements. We obtain an average emissivity index of 1.47 $\pm$ 0.17 across the complex, consistent with the results derived for the whole LMC by \citet{Planck_collabo_2011_MagellanicClouds} ($\sim$1.5) or for a strip in the LMC using \hersc\ Science Demonstration Phase data by \citet{Gordon2010}. We thus decide to fix the effective emissivity index to 1.5 and re-derive the temperature map. Fixing the emissivity index value prevents biases linked with {\it 1)} temperature-emissivity anti-correlation due to measurement uncertainties, {\it 2)} temperature mixing along the line of sight contributing to a bias in the derived ``effective emissivity index"  \citep[see][for discussions on these issues]{Juvela2012,Galametz2012}. \\

{\it Error estimates - }We perform Monte Carlo simulations in order to quantify the uncertainties on dust temperatures driven by flux errors of the resolved fluxes. Error maps are a combination in quadrature of the uncertainty maps derived during the data reduction (mapping, pointing) and uncertainties in the absolute calibration (15$\%$ for the SPIRE bands that are the most crucial for dust mass estimates), the latter being the dominant source of uncertainties. We generate 20 sets of modified constraints with fluxes randomly varying within their error bars and following a normal distribution around the nominal value and run the model for each set. Resolved uncertainties can thus derived using the standard deviations of each distribution. Since the absolute calibration is correlated for SPIRE bands, the three SPIRE measurements move consistently in each set. \\ 

\subsection{Realistic dust properties fitting}

We also apply the phenomenological SED fitting procedure of \citet{Galliano2011} to our 3.6 to 500 \mic\ data in order to derive the resolved total infrared (TIR) luminosities, dust properties and radiation field intensity estimates across the region. Their approach is similar to that of \citet{Draine2007}, except for the starlight intensity distribution. In both models, as suggested by \citet{Dale2001}, the distribution of starlight intensities per unit dust mass is approximated by a power-law (of index $\alpha$). In the \citet{Draine2007} models though, a significant fraction of the dust mass is assumed to be heated by starlight with a single intensity U$_{min}$ (``diffuse ISM" component). This model leads to better resolved fits (lower $\chi$$^2$ values) for nearby spirals \citep[see][]{Aniano2012} but does not seem to correctly reproduce LMC observations \citep{Galliano2011}. We choose the approach of \citet{Galliano2011} in order to allow flexibility of the model at submm wavelengths and account for possible variations of the effective emissivity or the presence of colder dust.  

We consider the dust size distribution and composition to be uniform across the complex and that sources of IR emission are old stars and dust (PAHs, carbon grains and silicates). The stellar contribution to the MIR is modelled using a library of stellar spectra previously synthesised using the stellar evolution code PEGASE \citep{Fioc_Rocca_1997}. The stellar population is assumed to have undergone an instantaneous burst 5 Gyr ago (initial solar metallicity, Salpeter Initial Mass Function). We do not have a MIR spectrum to properly constrain the PAH composition. We thus choose to fix the ionised PAH-to-neutral PAH ratio (f$_{PAH+}$), an additional parameter of the \citet{Galliano2011} model, to 0.5 in order to limit the number of free parameters of the model. \\

The free parameters of our model are thus: 
\begin{itemize}
\item the total mass of dust (M$_{dust}$), 
\item the PAH-to-dust mass ratio (f$_{PAH}$), 
\item the index describing the fraction of dust exposed to a given intensity ($\alpha$), 
\item the minimum heating intensity (U$_{min}$), 
\item the range of starlight intensities ($\Delta$U),
\item the mass of old stars (scaling of our stellar spectra).
\end{itemize}

We note that U=1 corresponds to the intensity of the solar neighbourhood (2.2$\times$10$^{-5}$ W~m$^{-2}$). For the rest of the paper, U$_{min}$+$\Delta$U will be referred as ``U$_{max}$". The PAH fraction f$_{PAH}$ is normalised to the average Galactic value \citep[4.6 $\%$;][]{Li_Draine_2001}. As pointed out in \citet{Galliano2011}, the mass of old stars is poorly constrained by the IRAC bands and is only used to subtract the stellar continuum to the MIR bands. No balance is made between this parameter and the starlight intensity. Errors on the physical parameters are derived using the same Monte Carlo technique described in Section 3.1. \\

{\it Amorphous Carbon to model carbon dust - }
Two different approaches are compared in \citet{Galliano2011} as far as the modelling of carbon dust is concerned: the ``standard model" that uses the \citet{Zubko2004} Galactic grain composition and graphite grains, or the ``AC model" that uses amorphous carbons \citep{Zubko1996} in lieu of graphite. Carbon constitutes a major fraction of the grains in galaxies, especially in the circumstellar dust of carbon stars. However, the exact form of carbon dust is still rather unclear. Many SED models use graphite to describe the interstellar carbon dust as first suggested by \citet{Mathis1977}. Graphite, indeed, has well-known physical properties, which makes it easy to model, and is in good agreement with the observed (Galactic) extinction and polarization \citep[][]{Draine_Lee_1984}. 
However, observational evidence challenges the graphite theory (at least the pure mono crystalline graphite theory), among which are {\it 1)} variations in the 2175 $\AA$ profile not explained by changes in graphite grain size \citep{Draine1993}, {\it 2)} a graphite broadband emission feature near 33 \mic\ less strong (weaker and narrower) than expected in global models \citep[e.g.][]{Draine2007}, {\it 3)} an erosion of carbon dust in shocks not reproducible in the case of graphite grains. Indeed, \citet{Welty2002} or \citet{Serra_Dias_Cano2008} show that graphite is not expected to survive as such in the ISM due to erosion and irradiation in shock waves and suggest that interstellar hydrogenated amorphous carbons could be the most probable form of carbon material, their erosion being more efficient than graphite in shocks. We will also see further in the paper that using the ``standard model" leads to dust masses inconsistent with the dust-to-gas mass ratio expected for the complex. The use of AC to model carbon dust generally leads to a decrease of the mass of large grains compared to that required by the ``standard model" to account for the same emission (see Section 4.5.1 for explanations). Considering the average emissivity index of our region ($\beta$$\sim$1.5), we favour the ``AC" approach to model the resolved SEDs across the complex.

%------------------------------------------------------------------------------------------------------------------------------------------
\begin{figure*}
    \centering
     \includegraphics[width=18cm]{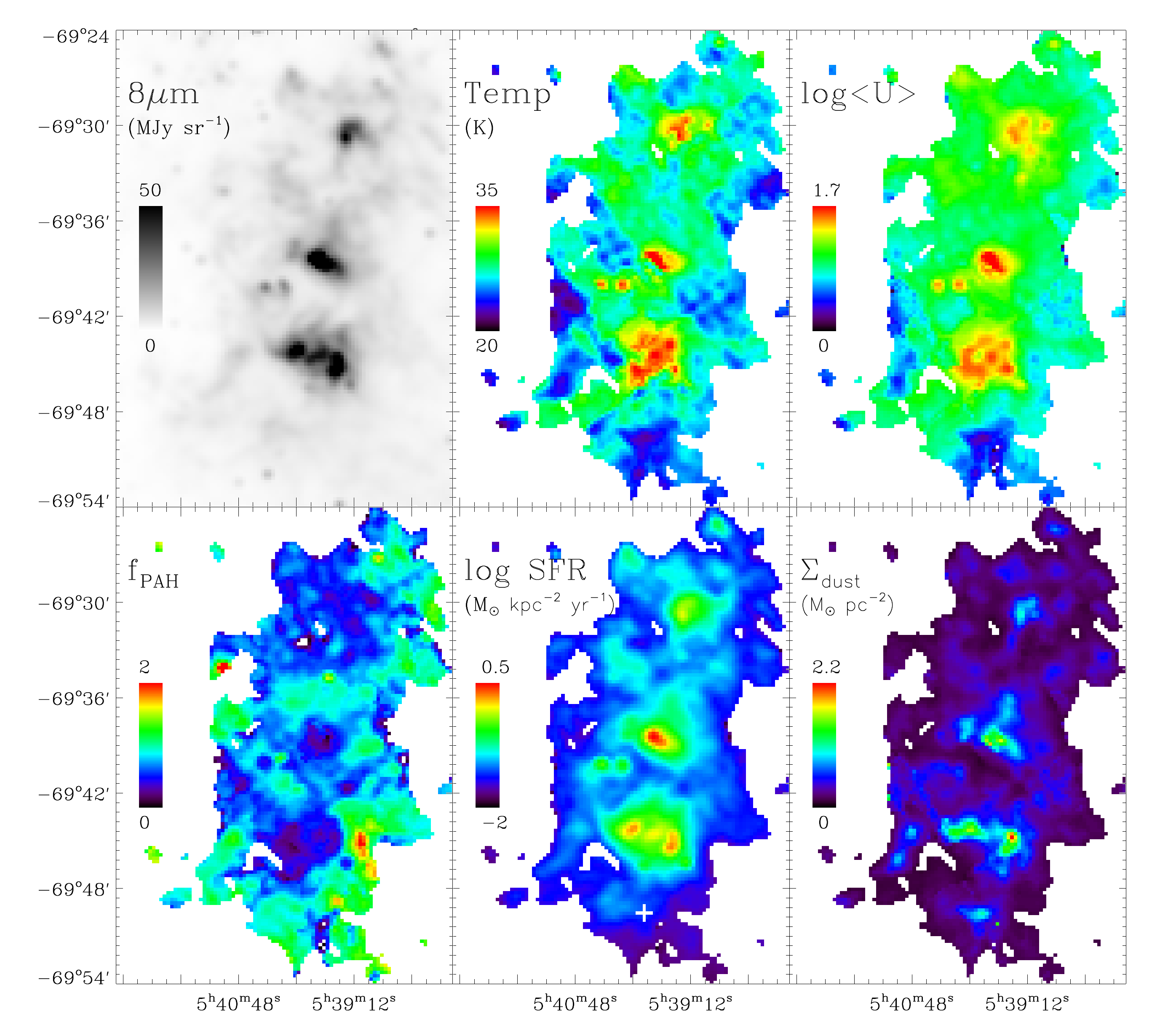}     \\    
     \caption{Results of our resolved SED fittings: {\it Upper left:} 8 \mic\ map convolved to the SPIRE 500 \mic\ resolution (units of MJy~sr$^{-1}$). {\it Upper middle:} Cold dust temperature (units in Kelvins). {\it Upper right:} Mean starlight heating intensity (U=1 corresponds to the intensity of the solar neighbourhood). {\it Lower left:} PAH-to-total dust mass ratio normalised to the Galactic value (4.6$\%$). {\it Lower middle:} Star formation rate (units in \msun~kpc$^{-2}$~yr$^{-1}$ - log scale). The white cross indicates the N159S region. {\it Lower right:} Dust surface density (in \msun~pc$^{-2}$). For all images, North is up, east is left and coordinates (RA, Dec) are given in equinox J2000.}.
    \label{ModelResults}
\end{figure*}
%------------------------------------------------------------------------------------------------------------------------------------------

\section{Results and analysis}

In this section, we describe the various maps obtained directly or indirectly from our resolved SED models and analyse how their distributions correlate.  Figure~\ref{ModelResults} shows the maps of the parameters we derive (temperature and mean radiation field intensity, PAH fraction, star formation rate and dust mass surface density). The 8\mic\ image is shown on the upper left panel for comparison. In Section 4.4, we also investigate how submm colours evolve with the ISM conditions (heating sources, radiation field intensities). Finally, we analyse the dust surface density map and how it relates to the distribution of the different gas phases in the complex, and more particularly in the N159 region in Section 4.5. 

\subsection{Temperatures and mean radiation field intensities}

We show the final cold temperature (i.e. temperature of the cold MBB) map obtained using our resolved two-MBB fitting technique in Fig.~\ref{ModelResults} (upper middle panel). The temperatures peak at the locations of star-forming regions where dust is expected to be warmer. N160 shows the highest cold dust temperature, with a maximum of $\sim$40K while the median temperature of the N158-N159-N160 complex is 26.9$^{\pm2.3}$K (28.2 if we restrict the analysis to ISM elements with a 3-$\sigma$ detection in the SPIRE bands). Based on observations made with the TopHat telescope combined with DIRBE data (100, 140, and 240 \mic, 42\arcmin), \citet{Aguirre2003} derived temperatures of 25$^{\pm1.8}$K and 26.2$^{\pm2.3}$K for the LMC and 30Dor, using $\beta$= 1.33$^{\pm0.07}$ and 1.5$^{\pm0.08}$, respectively. From \spitz\ data (thus no data above 160 \mic) and the same emissivity values, \citet{Bernard2008} obtained a colder temperature of 21.4 K for the LMC and 23 K for the 30 Dor region. Our values are thus close to the values derived for the 30Dor region by \citet{Aguirre2003}. We note however that different methods were used to derive the temperatures, making a direct comparison difficult. The N159S region is finally the coldest region of the whole complex, with a temperature $\sim$22K. This supports the argument that massive stars may still be at infancy and deeply embedded in the circumstellar material in this particular region.

In order to investigate more precisely the distribution of radiation field intensities across the region, we derive a map of the mass-weighted mean starlight heating intensity $<$U$>$ \citep{Draine_Li_2007} that develops into:
 
 {\large 
 \begin{equation}
<U>=\left \{
\begin{array}{ll}
\vspace{7pt}
{\large \frac{\Delta U}{ln~(~1~+~\Delta U/U_{min}~)}} &if~\alpha = 1 \\
\vspace{7pt}
U_{min}~\frac{ln~(~1+~\Delta U/U_{min}~)}{\Delta U~/~(U_{min}~+~\Delta U)}  &if~\alpha = 2 \\
\vspace{7pt} 
 \frac{1-\alpha}{2-\alpha}\frac{(U_{min}+\Delta U)^{2-\alpha}~-~U_{min}^{2-\alpha}}{(U_{min}+\Delta U)^{1-\alpha}~-~U_{min}^{1-\alpha}}& if~\alpha \ne 1~\&~\alpha \ne 2 \\
\end{array} \right .
\end{equation}
}

with $\Delta$U the range of starlight intensities, U$_{min}$ the minimum heating intensity and $\alpha$ the index describing the fraction of dust exposed to a given intensity, all three obtained from the resolved \citet{Galliano2011} modelling procedure. U=1 corresponds to the intensity of the solar neighbourhood (2.2$\times$10$^{-5}$ W~m$^{-2}$). \\

We show the mean starlight heating intensity map in Fig.~\ref{ModelResults} (upper right panel). Since dust grains are heated by the interstellar radiation field, we obtain a similar distribution between dust temperatures and radiation field intensities. Maxima in $<$U$>$ appear close to H\,{\sc ii} regions where the starlight heating the dust is expected to be much more intense than the overall starlight illuminating the complex. The averaged intensity in our ISM elements is $<$U$>$=8.6. This average increases to 11.7 when we restrict the calculation for resolved ISM elements that have a 3-$\sigma$ detection in the SPIRE bands. We note that \citet{Meixner2010} derived an average value $<$U$>$=7.6 for a strip across the LMC (so including more quiescent regions) using the same model as that applied in our study. Finally, N159S shows the lowest $<$U$>$ ($\sim$1.8) of the complex. This is consistent with the cold temperatures we observe in this region. This region thus appears to be shielded from the strong starlight heating the N159 H\,{\sc ii} region.

\subsection{PAH fraction} 

Figure~\ref{ModelResults} (lower left panel) shows how the PAH-to-total dust mass ratio f$_{PAH}$ is distributed across the star-forming complex.  We remind the reader that in the modelling, f$_{PAH}$ is normalised to the Galactic value f$_{PAH,MW}$=4.6$\%$. The average value of f$_{PAH}$ across the region is driven by the ISM outside the H\,{\sc ii} regions themselves (median f$_{PAH}$=0.7$^{\pm0.3}$). However, we observe variations in the PAH fraction. Peaks of the f$_{PAH}$ (in red) are located in low-surface brightnesses regions and are probably artefacts due to the very low dust mass  of those regions. We observe a strong decrease of f$_{PAH}$ in the three star-forming regions as well as low PAH fractions in two compact knots located at the south east side of N160. All these regions correspond to peaks in the radiation field intensities (Fig.~\ref{ModelResults}, $<$U$>$ panel) as well as strong radio emission (as we will see later in Section 5). This depression of the PAH fraction by intense radiation fields was also studied in detail by \citet{Sandstrom2010_2} in the SMC. These results are consistent with the fact that PAHs are known to be destroyed in strong H\,{\sc ii} regions \citep{Madden2006,LeBouteiller2007}. PAHs are also smaller on average in low-metallicity galaxies, which could also favour their destruction \citep{Sandstrom2012}. Low-metallicity environments often show harder interstellar radiation fields extending over larger size scales. This effect is attributed to the hardness of the radiation field and/or to a lower dust attenuation in low-metallicity ISM which consequently leads to a larger mean free path length of the ionising photons. Moreover, the fragmented structure and clumpiness of the ISM in low-metallicity galaxies allows UV light to penetrate deeper in metal-poor molecular clouds, favouring grain destruction. We note that should the radiation field be harder than the interstellar radiation field of \citet{Mathis1983} used in our model, the PAH fraction we estimate could be over-estimated.

\citet{Dwek2005} and \citet{Galliano_Dwek_Chanial_2008} have suggested that the low fraction of PAHs in metal-poor environments could also be due to a delayed injection of carbon dust into the ISM by AGB stars in the final phase of their evolution. Using near-IR JHK photometry in the N159 nebula, \citet{Meynadier2004} distinguished a young ($\sim$3 Myr) and an old ($\sim$1-10 Gyr) stellar population, even if they could not exclude that both populations might be spatially unrelated. Based on the absence of red supergiants in N159 (stars that should normally be observed in the cluster if the present stellar population was at least 10 Myr old), \citet{Jones2005} concluded that most of the star formation must have taken place recently and that N159 might not be older than 1-2 Myr. A sequential cluster formation from N160 to N159S has been observed by \citet{Nakajima2005}, using observations of Herbig Ae/Be clusters. This supports a slightly longer star formation activity for the N160 region ($\sim$10-30 Myr). Both regions are nevertheless quite young compared to the time necessary for carbon dust to be injected in the ISM by the stars and thus participate in the formation of PAHs in the clusters. We finally note that recent studies suggest than the rate of AGB-produced PAHs compared to the rate of PAHs formed in the ISM may also change with metallicity \citep{Sandstrom2012}.\\

In conclusion, although the obvious anti-correlation between f$_{PAH}$ and the radiation field favours the destruction of PAHs as a straight forward explanation for the low PAH fraction we observe, the young age of the clusters could be an additional explanation for the absence of PAH in our three H\,{\sc ii} regions.

\subsection{Star Formation rate}

Since the total bolometric IR luminosity traces the emission of stellar populations enshrouded in dust cocoons, this quantity is thus very often used to quantify the star formation obscured by dust \citep{Perez-Gonzalez_2006,Kennicutt2009}. We first obtain the total IR luminosity map of the complex by integrating the resolved SEDs in a $\nu$-f$_{\nu}$ space:

\begin{equation}
L_{IR} = \int_{8~\mu m}^{1100~\mu m}L_{\nu}~d\nu
\label{eq2}
\end{equation}

In the SED fitting technique we use, we model the stellar contribution to short wavelengths as a separate component. This contribution is subtracted before performing the integration to only take the thermal emission of dust into account in the calculation. We use the IR-based calibration of \citet{Kennicutt1998} to derive a star formation rate map of the region: 

\begin{equation}
\frac{SFR}{M_{\odot}~yr^{-1}}=\frac{L_{FIR}}{5.8 \times 10^9 L_{\odot}}
\label{eq3}
\end{equation} 

 \citet{Jones2005} estimated the integrated luminosity of the N159 complex (integration of the SED up to 100\mic\ combining \spitz\ and KAO data). They found that the integrated luminosity was consistent with the observed radio emission, assuming that the H\,{\sc ii} region was a cluster with a normal Initial Mass Function (IMF). As mentioned in \citet{Indebetouw2008}, Eq.~\ref{eq3} was calibrated for large extragalactic sources. It may be valid for the full star-forming regions such as N159, but the SFR calibration may not be accurate when applied to some of our resolved ISM elements, namely those not properly covering the stellar IMF. \\

We show the SFR map of the region in units of \msun~yr$^{-1}$~kpc$^{-2}$ (log scale) in Fig.~\ref{ModelResults} (lower middle panel). We find a median SFR of 0.06 \msun~yr$^{-1}$~kpc$^{-2}$ across our modeled region. This average increases to 0.17 \msun~yr$^{-1}$~kpc$^{-2}$ when we restrict the calculation for resolved ISM elements that have a 3-$\sigma$ detection in the SPIRE bands. This is consistent with the average SFR obtained by \citet{Indebetouw2008} using the H$_2$ surface densities and the Schmitt-Kennicutt law (0.14 \msun~yr$^{-1}$~kpc$^{-2}$ for N159/N160 and 0.11 \msun~yr$^{-1}$~kpc$^{-2}$ for the southern molecular ridge to whom the complex belongs to). They find that their SFR estimate for N159/N160 corresponds to that derived using the \citet{Calzetti2007} formula (i.e. a calibration from H$\alpha$ and 24 \mic\ luminosities) within uncertainties. We note that the SFR we derive varies significantly across the region. Peaks of SFR correspond to our three star-forming regions, with the N160 centre showing the highest star formation rate of the complex. We indicate the N159S region by a white cross in Fig.~\ref{ModelResults} (SFR panel). The region does not show massive on-going star formation.

%----------------------------------------------------------------------------------------------------------------------------
\begin{figure*}
    \centering
        \vspace{-10pt}
    \begin{tabular}{c}
    \vspace{-10pt}
    \hspace{-10pt}
\includegraphics[width=18cm ]{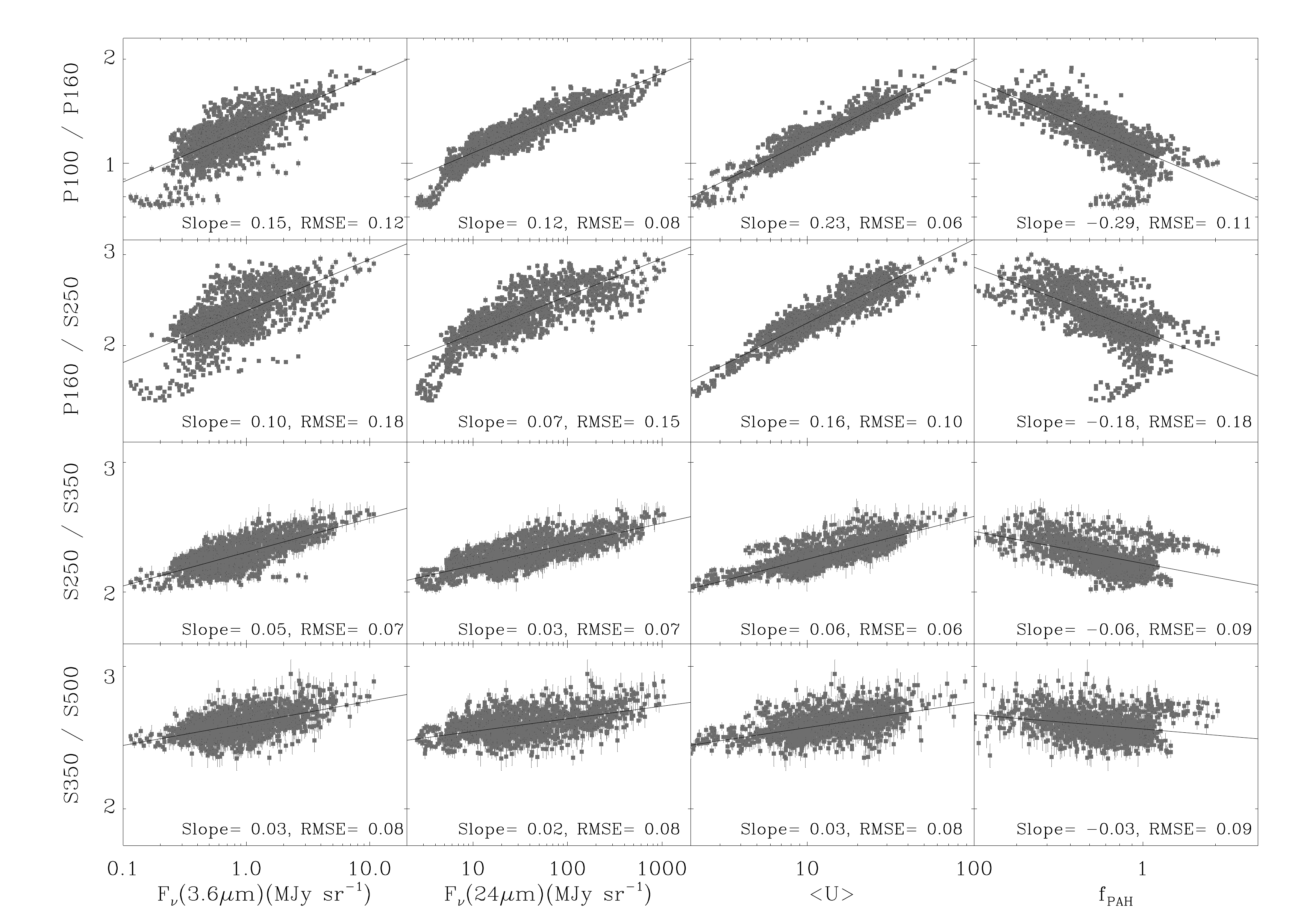}\\
\end{tabular}
    \caption{Comparison of PACS and/or SPIRE surface brightness ratios to the 3.6 \mic\ surface brightnesses, the 24 \mic\ surface brightnesses, the mean stellar density $<$U$>$ or the PAH fraction f$_{PAH}$. Black lines show the best fit lines and we provide the slope and the RMSE at the bottom of each panel. }
    \label{SPIRE_colour_1}
\end{figure*}
 %----------------------------------------------------------------------------------------------------------------------------

%----------------------------------------------------------------------------------------------------------------------------
\begin{figure*}
    \centering
        \vspace{-10pt}
    \begin{tabular}{c}
    \vspace{-10pt}
\includegraphics[width=13cm ]{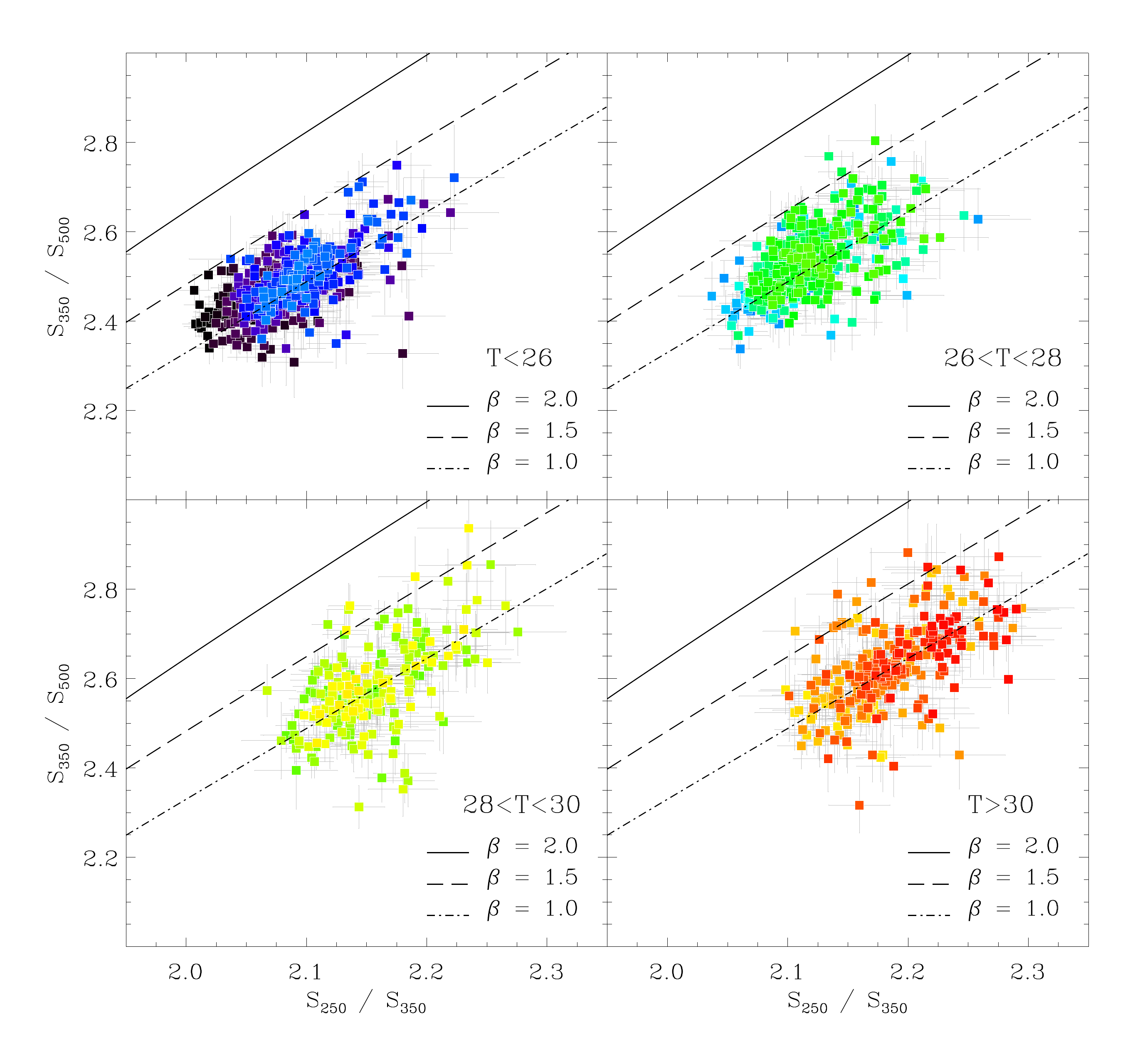}\\
\end{tabular}
\vspace{-10pt}
    \caption{Colour-colour plot of the star-forming complex: 350/500 colours are plotted against the 250/350 colour. Squares represent the resolved ISM elements of the star-forming complex (pixel size = 14\arcsec\ i.e. $\sim$4pc for the LMC) coded with temperature (temperatures taken from the map presented in Fig.~\ref{ModelResults}). We separate the resolved elements in 4 panels (T$<$26, 26$<$T$<$28, 28$<$T$<$30 and T$>$30). For each panel, elements are binned 3-by-3 for clarity. The lines show the colours obtained from a single temperature modified blackbody with an emissivity index $\beta$$_{singleBB}$ of 2, 1.5 and 1, from cold temperatures to the left to hot temperature to the right. }
    \label{SPIRE_colour_2}
\end{figure*}
 %----------------------------------------------------------------------------------------------------------------------------

 \subsection{Submillimeter colours}

\citet{Bendo2011} studied \hersc\ colours (i.e. PACS or SPIRE surface brightness ratios) in three nearby galaxies (M81, M83 and NGC~2403) and found that the 70/160 colour was correlated with tracers of the star formation activity (thus that warm dust is primarily heated by the young stellar populations) while SPIRE ratios (250/350 or 350/500) were more strongly correlated with the galacto-centric radius of the galaxies. They concluded that cold dust could be primarily heated by the total stellar population (young stars + old stars). \citet{Boquien2010} and \citet{Galametz2010} studied similar correlations in M33 and NGC~6822 respectively and showed that these ratios were also strongly correlated with the star formation, traced by the 24 \mic\ emission for instance. In Section 4.4.1, we thus study the dependence of FIR colours on various parameters, including the mean radiation field intensities $<$U$>$ and the PAH fraction just derived from our modelling.
Moreover, if FIR colours are often used as good proxies for dust temperatures, colour-colour relations can, on the other hand, provide additional information about the dust emissivity wavelength dependence \citep{Nagata2002,Hibi2006,Hirashita2009_2}. In Section 4.4.2, we analyse SPIRE colours as possible indicators of the cold dust emitting properties. 

\subsubsection{Dependencies}

In order to investigate the drivers of submm colours variations in the N158-N159-N160 complex, we compare in Fig.~\ref{SPIRE_colour_1} the PACS and SPIRE surface brightness ratios 100/160, 160/250, 250/350 and 350/500 with :

\begin{enumerate}

\item[1)] the 3.6 \mic\ surface brightness: dominated by emission from the old stellar populations, it is known as a good proxy of the stellar mass \citep[see][for recent studies]{Zhu2010,Eskew2012},

\item[2)] the 24 \mic\ hot dust emission, often used as a good calibrator for star formation \citep[][among others]{Calzetti2007,Rieke2009},

\item[3)] the mean starlight heating intensity $<$U$>$. Values of U have been normalised to that of the Milky Way),

\item[4)] the PAH fraction f$_{PAH}$. 
\end{enumerate}

For this section as well as the following, we restrict ourselves to pixels with a 3-$\sigma$ detection in the three SPIRE bands. We note that part of the N159S region matches this signal-to-noise criteria. In Fig.~\ref{SPIRE_colour_1}, we compute the linear regressions from each comparison using an ordinary Least-Squares regression (function {\it linfit.pro} of the Astrolibrary of IDL). On each plot, we provide the slope of the relation as well as the root-mean-square error defined as:

\begin{equation}
RMSE=\sqrt{\frac{\Sigma~(O_i-E_i)^2}{n}}
\end{equation}

\noindent with O$_i$ the observed ratios (or temperature colours), E$_i$ the ratios estimated from our fitting relations and n the number of ISM elements.\\

We observe correlations in all cases, with a very tight relation between the 100-to-160 and 160-to-250 flux density ratios and the mean starlight intensity $<$U$>$ (lower RMSE for each line). We know that dust grains are heated by the interstellar radiation field and re-emit this energy in the infrared regime so correlation between FIR colours and the radiation field intensity are expected. FIR colours are often used as good proxies of the average dust temperatures, even if FIR ratios using shorter wavelengths than 100 \mic\ (70-to-100 ratio for instance) can be potentially biased by emission from stochastically heated dust grains \citep{Dale2001,Bernard2008}. Contrary to ratios at shorter wavelengths that show stronger correlations with the 24 \mic\ surface brightnesses, the SPIRE-only ratios follow the 3.6 and 24 \mic\ surface brightnesses equally well. We finally note that the N159S region shows the lowest submm surface brightness ratios of the whole complex, with surface brightness ratios of $\sim$0.75 for 100/160, $\sim$1.58 for 160/250, $\sim$2 for 250/350 and $\sim$2.35 for 350/500.

\subsubsection{SPIRE colour-colour diagrams}

Figure~\ref{SPIRE_colour_2} shows a SPIRE colour-colour diagram of the star-forming complex. We plot the 350-to-500 \mic\ flux density ratios (hereafter 350/500 colour) of our resolved elements as a function of the 250-to-350 \mic\ flux density ratios (hereafter 250/350 colour). In order to investigate variations in the submm colours with temperature, we separate the points in 4 panels (T$<$26, 26$<$T$<$28, 28$<$T$<$30 and T$>$30) and code them with temperature, from cold in dark purple to hot in red colours. We use the resolved temperatures T$_c$ derived using our two-modified blackbody model. The lines show modelled colours obtained from a single temperature modified blackbody with an emissivity index $\beta$$_{singleBB}$ of 2, 1.5 or 1. 

We observe that SPIRE colours are well correlated with each other. Resolved elements with T$>$30 (fourth panel) have, on average, slightly lower 350/500 colours for a given 250/350 colour, so a ``flatter" submm slope. This trend agrees with the anti-correlation observed between the emissivity index and the temperature \citep{Dupac2003,Shetty2009} but could also result from temperature mixing effects along the line of sight in these elements. Given our error bars, we, however, consider this trend as marginal. We note that the ISM elements showing very low 250/350 colours (black squares) belong to N159S. 

As suggested by Fig.~\ref{SPIRE_colour_2}, a single modified blackbody model with $\beta$$_{singleBB}$=2 cannot reproduce the colours we observe. In fact, many resolved elements reside around the $\beta$=1 model, thus at the very low end of the $\beta$ values quoted in the literature (1$<$$\beta$$<$2.5). This cautions the use of single-temperature fits to derive the intrinsic emissivity of the dust grains, those models providing in reality an ``effective" emissivity index resulting from the combination of various dust populations at different temperatures.

\subsection{Dust and gas masses across the complex}

\subsubsection{Dust masses}

We show the dust surface density map $\Sigma$$_{dust}$ obtained with the ``AC model" in Fig.~\ref{ModelResults} (lower right panel). The distribution of the dust surface density follows the distribution of the FIR/submm emission. We particularly point at the elongated structure of N159 correlated with the SPIRE and 870 \mic\ emission. The total dust mass of the region we modeled is $\sim$2.1$\times$10$^{4}$ \msun, with 8.5$\times$10$^{3}$ \msun\ within resolved elements fulfilling the 3-$\sigma$ SPIRE-band detection\footnote{We note that estimating this total dust mass on a resolved scale (resolved masses added together) allows us to probe the various dust components in ISM elements with low-dust mass surface densities as well as in dense regions. Contrary to dust mass estimates derived from a global SED modelling of the region, we can avoid potential biases (namely a dust mass underestimation) caused by a poorer resolution \citep[c.f. discussion about biases caused by non-linearities in SED models in][]{Galliano2011}.}. A significant amount of dust is present in N159S, a region whose emission starts to be detected at 160 \mic\ with PACS (Fig.~\ref{Spitzer_Herschel_LABOCA_maps}), thus a significant reservoir of cold dust. We refer to Section 5 for an individual SED modelling of N158, N159, N159S and N160 that allows us to derive individual dust masses for these H\,{\sc ii} regions.

The resolved dust masses we estimate are 2.8 times higher if we use the ``standard model" of \citet{Galliano2011}, i.e. graphite grains to model carbon dust. This is consistent with the statistical distribution of the resolved dust mass ratio between the ``AC model" and the ``standard model" they find in a strip across the LMC (median of the distribution at $\sim$2.6). In fact, the opacity profile of AC is higher than graphite for $\lambda$$>$5\mic. This implies that for a fixed starlight intensity, AC will have a lower temperature than graphite, thus lead to more mass. However, AC has a flatter submm slope (more emissivity in the submm regime\footnote{c.f. further discussions about opacities in \citet{Galliano2011}, Appendix A}), so we are able to fit the same observational constraints with slightly hotter grains but less mass. Indeed, the ``AC model" invokes higher starlight intensities (median of $<$U$>$ across the complex: 8.6 in the ``AC case", 4.1 in the ``standard model"), and thus less dust than graphite to account for the same submm emission. We finally note that the errors on resolved dust masses are of the order of $\sim$50$\%$. This will introduce non-negligible uncertainties in our gas-to-dust mass ratio (G/D) estimates (c.f. Section 4.5.4).

\subsubsection{Comparison with the H\,{\sc i} distribution}

High-resolution maps of the atomic gas reservoir have been built for the complete LMC. The H\,{\sc i} data cube used in this section \citep[from][]{Kim2003} combines data from ATCA, the Australian Telescope Compact Array \citep{Kim2003} and the Parkes single-dish telescope \citep{Staveley2003}. The individual channel maps have a velocity resolution of 1.649 km s$^{-1}$ and the H\,{\sc i} map was obtained integrating the cube over the range 190 $<$ v$_{hel}$ $<$ 386 km s$^{-1}$, with a resolution of 1\arcmin. We refer to \citet{Kim2003} and \citet{Bernard2008} for further details on the LMC integrated map.

\begin{figure}
    \centering
    \begin{tabular}{c}
\hspace{-10pt}\includegraphics[width=9.5cm]{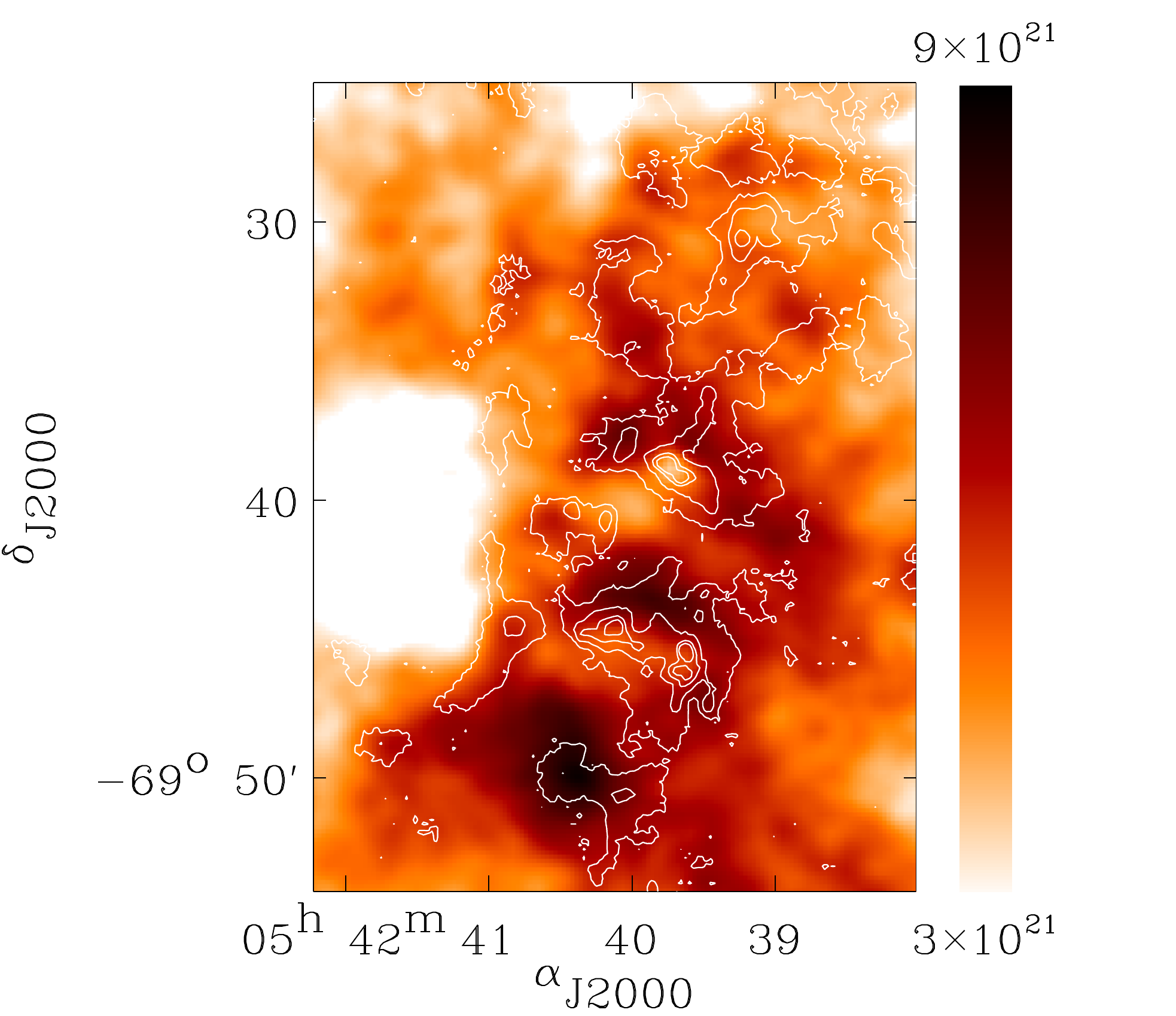} \\
\hspace{-10pt}\includegraphics[width=8.5cm]{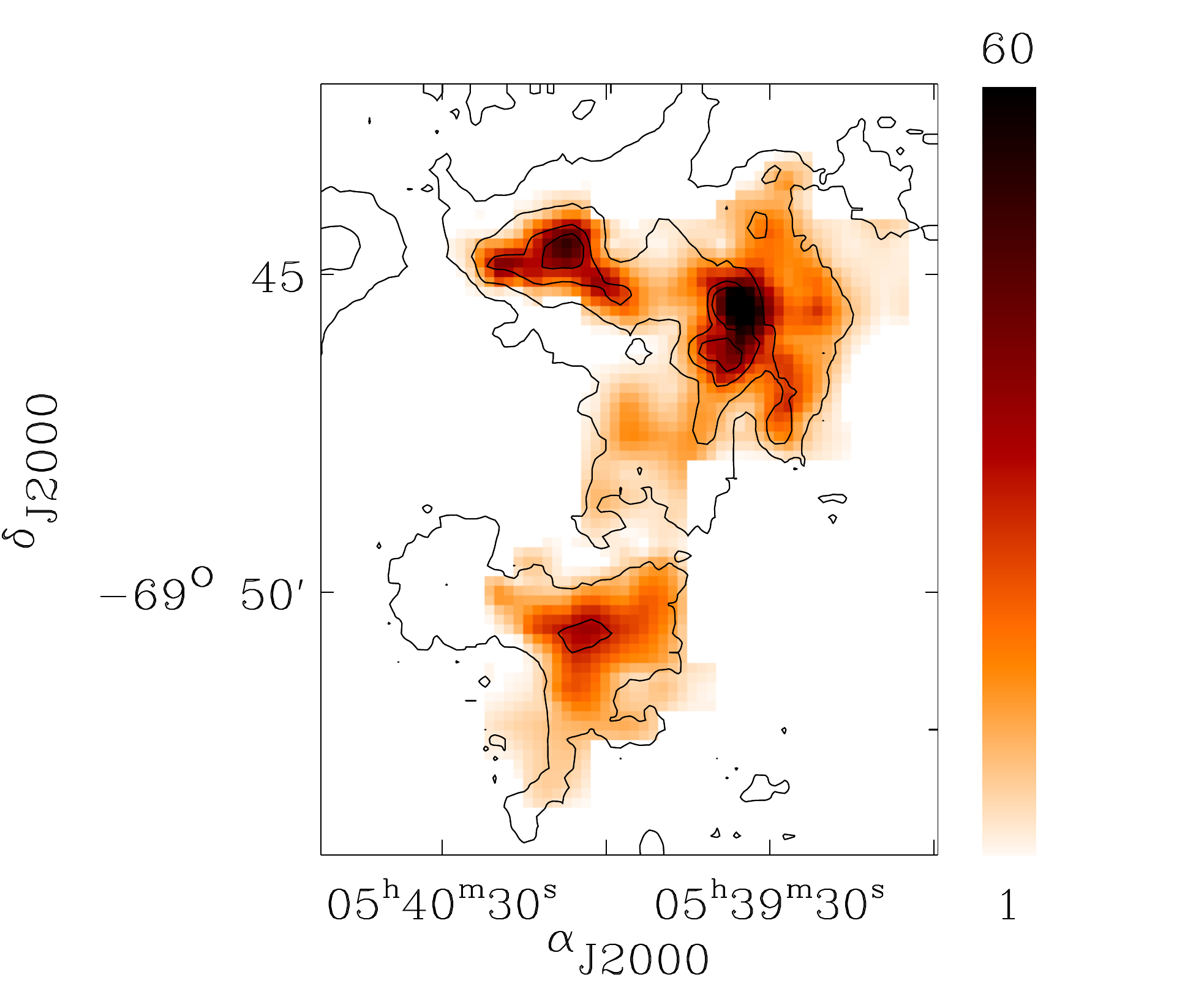}\\
\end{tabular}
    \caption{{\it Top panel:} The N158-N159-N160 complex observed in H\,{\sc i} (in units of atom cm$^{-2}$). The H\,{\sc i} map is a combination from ATCA \citep{Kim2003} and Parkes \citep{Staveley2003} observations. 870 \mic\ contours are overlaid (0.02, 0.1, 0.25 and 0.4 Jy~beam$^{-1}$). {\it Bottom panel:} ASTE CO(3-2) observation of N159 (in units of K km s$^{-1}$). Contours: same than top. For both images, North is up, east is left.}
    \label{LABOCA_HI_CO}
\end{figure}

Figure~\ref{LABOCA_HI_CO} (top) shows the H\,{\sc i} map of the N158-N159-N160 complex, with 870 \mic\ contours overlaid. The distribution of H\,{\sc i} does not match that of the emission in the FIR (both \spitz\ and \hersc) bands across the region. A clear offset is for instance observed between N159 and the closest H\,{\sc i} peak located further north ($\sim$1.5\arcmin). A closer look at the regions indicates that the peaks of the 870 \mic\ emission of the three regions N158, N160 and N159, as well as the PACS and SPIRE peaks, reside in fact in H\,{\sc i} holes. This could be a signature of the H\,{\sc i} to H$_2$ transition already observed in LMC or Galactic clouds \citep[][Tatton et al. in prep, among others]{Wong2009,Lee2012}, with H\,{\sc i} being converted into H$_2$ either by thermal or gravitational instabilities or by shock compressions.
Contrary to the other regions of the complex, the southern H\,{\sc i} peak does spatially coincide with a peak in the 8 \mic\ emission (see Fig.~\ref{IRAC_LABOCA} bottom panel for comparison between the 8 and the 870 \mic\ emission in N159S). This peak is located $\sim$ 30 pc east of N159S.

%----------------------------------------------------------------------------------------------------------------------------
%The H\,{\sc i} contours are 5 $\times$ 10$^{21}$, 6 $\times$ 10$^{21}$ and 7 $\times$ 10$^{21}$ atom cm$^{-2}$. 
%The CO contours are 8, 16, 32 and 42 K km s$^{-1}$.
 %----------------------------------------------------------------------------------------------------------------------------
 
 %----------------------------------------------------------------------------------------------------------------------------
\begin{figure*}
    \centering
    \begin{tabular}{m{7cm}m{9cm}}
\includegraphics[height=7cm]{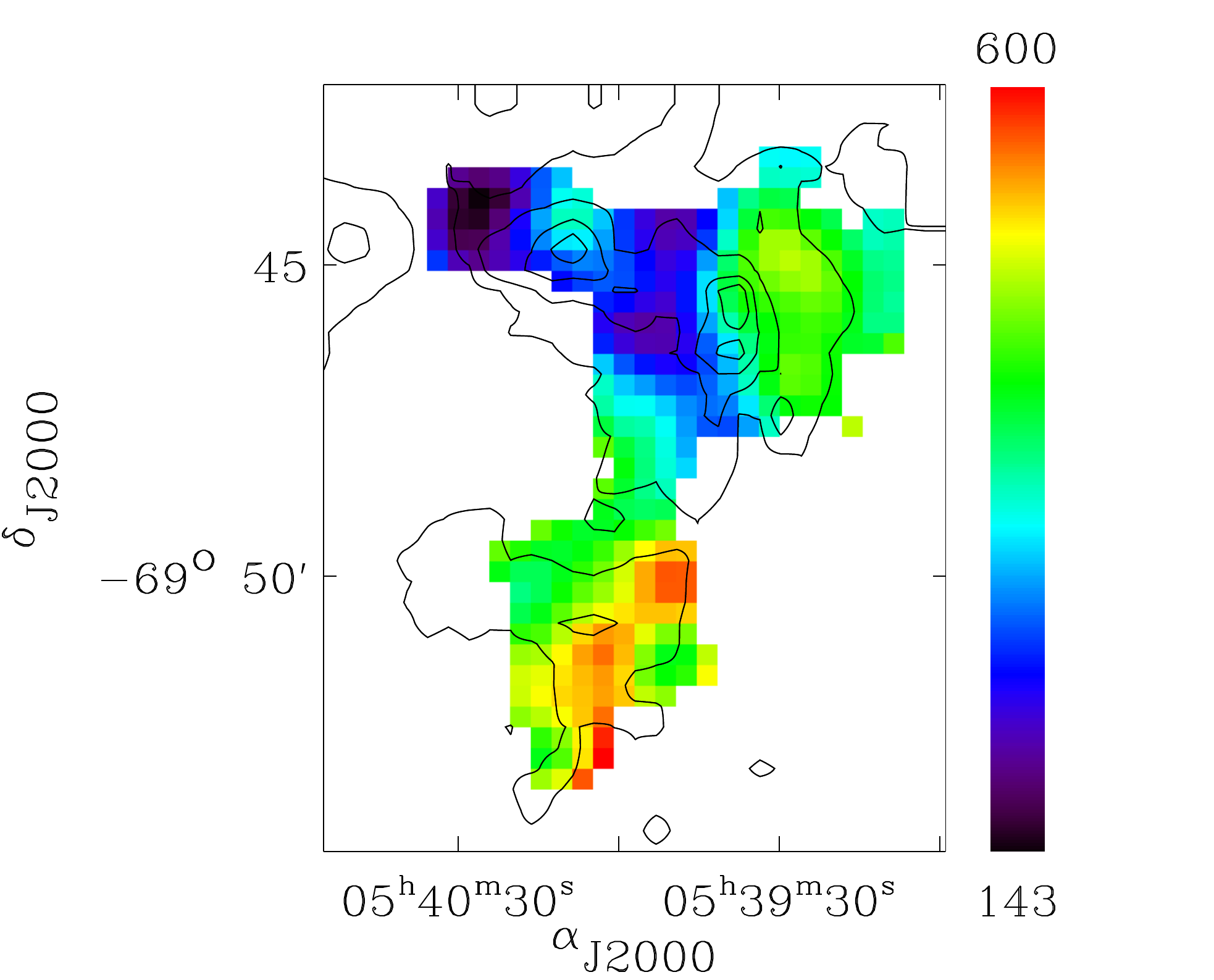} & \includegraphics[height=7.5cm]{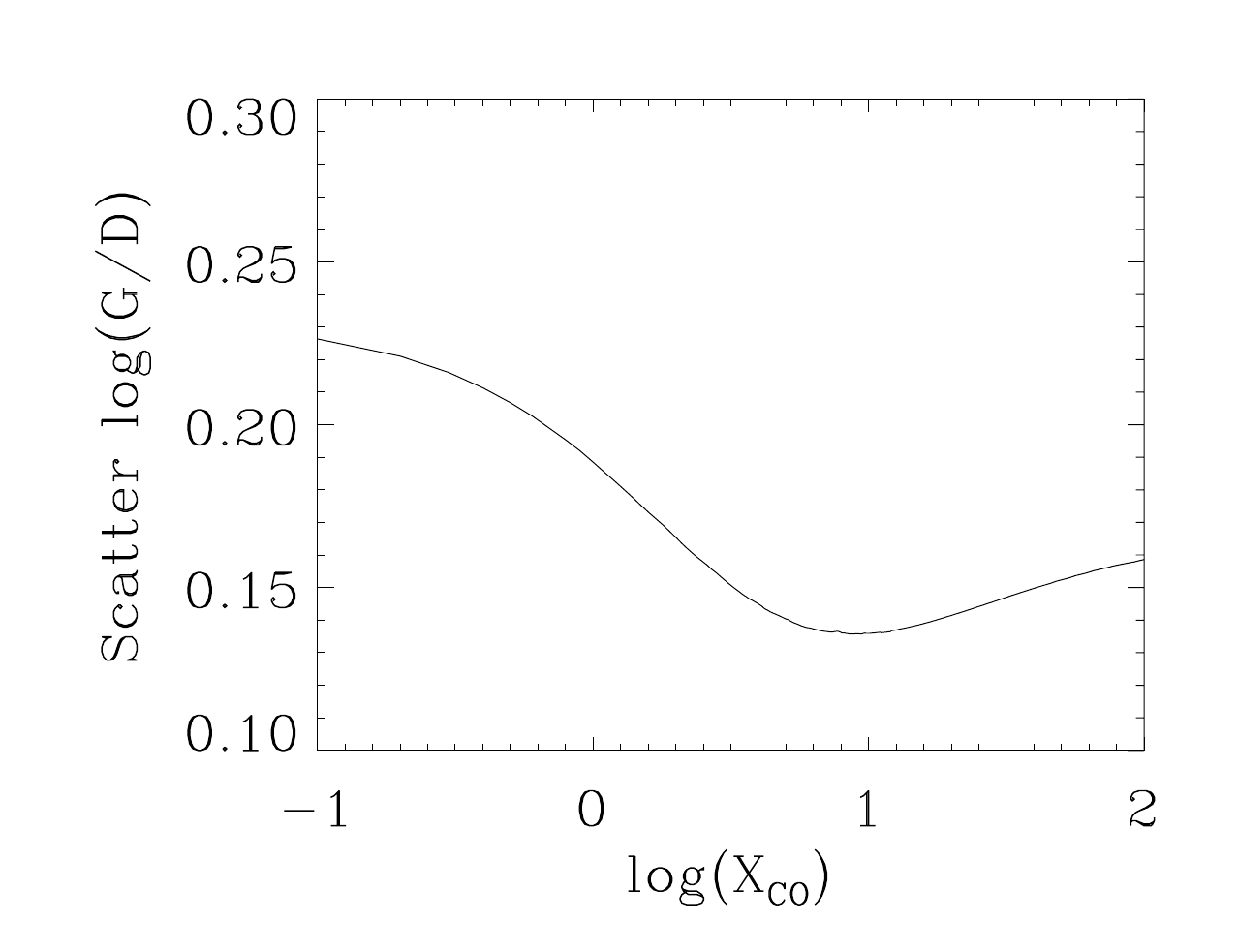}  \\
\end{tabular}
    \caption{{\it Left:} Map of the gas-to-dust mass ratio of the N159 region derived using X$_{CO}$=7 $\times$ 10$^{20}$ H$_2$ cm$^{-2}$ (K km s$^{-1}$)$^{-1}$ to convert CO observations in H$_2$ mass. 870 \mic\ contours are overlaid (0.02, 0.1, 0.25 and 0.4 Jy~beam$^{-1}$). {\it Right:} Scatter in the logarithm of G/D across the region as a function of the X$_{CO}$ factor used to convert CO observations in H$_2$ mass (in units of \msun~pc$^{-2}$(K km s$^{-1}$)$^{-1}$). We use the minimization technique described by \citet{Sandstrom2012_2}. The scatter is minimized when X$_{CO}$ = 8.6 \msun~pc$^{-2}$ (K km s$^{-1}$)$^{-1}$.}
    \label{DGR}
\end{figure*}
 %----------------------------------------------------------------------------------------------------------------------------

\subsubsection{CO in N159 and N159S}   
   
In this section, we focus more particularly in the N159 region to study how the dust budget and the dust properties are related to the molecular gas. Several CO observations of the N159 nebula have been carried out so far. \citet{Johansson1998} observed $^{12}$CO(J=1-0) with the SEST (Swedish-ESO Submillimeter Telescope). They obtained maps with a resolution of 45\arcsec, corresponding to $\sim$11pc for N159. \citet{Bolatto2000} also mapped the N158~/~N159~/~N160 complex in four transitions: $^{13}$CO(1-0), $^{12}$CO(2-1), $^{12}$CO(4-3), [C{\sc i}]($^3$P$_1$~$\rightarrow$~$^3$P$_0$) also with SEST and the 1.7m AST/RO (Antartic Submillimeter Telescope and Remote Observatory).
A spatially resolved survey of the LMC was also performed with the NANTEN 4-m telescope in the 2.6 mm carbon monoxide emission with a 40 pc resolution by \citet{Fukui1999,Fukui2008}, permitting a study of the giant molecular clouds at global scales. Several observations of the CO(4-3) transition and of the CO(3-2) transition have been carried out in selected regions of the LMC. These transitions have been observed in the N158~/~N159~/~N160 complex by \citet{Mizuno2010} ($^{12}$CO(J=4-3) and $^{13}$CO(J=3-2) and \citet{Minamidani2008} with NANTEN2 (beam size at 490 GHz: 18\arcsec) and ASTE (Atacama Submillimeter Telescope Experiment, beam size at 350 GHz: 22\arcsec-23\arcsec). Finally, the Magellanic Mopra Assessment (MAGMA) survey has recently completed a high angular resolution $^{12}$CO (1-0) mapping survey of giant molecular clouds across the LMC. The beam size of the instrument is 33\arcsec, which corresponds to a spatial resolution of $\sim$8 pc at the distance of the LMC. 

The two giant molecular clouds of N159 (N159E and N159W) as well as its southern region (N159S) are thus already mapped at high resolution in many transitions. Figure~\ref{LABOCA_HI_CO} shows the N159 region observed by ASTE CO(3-2) with 870 \mic\ contours overlaid. N159W is the strongest region and the eastern region breaks into 3 independent peaks. Compared to the other giant molecular clouds, the N159S does not show any heating source and little star formation activity. \citet{Bolatto2000} also note the detection of diffuse low level emission in CO(2-1). 

N159E and N159W are associated with embedded star clusters with temperatures that were determined to be 70-80K while N159~S shows a nearly uniform molecular temperature estimated to be $\sim$30K by \citet{Mizuno2010}. They also note that the molecular peak of the N159E region is showing an elongated structure that is very well resolved at 870 \mic\ with \lab. A slight shift appears between the peak of $^{12}$CO(4-3) in N159W and the 8 and 24 \mic\ peaks. The bright peak of the \lab\ emission in N159W does not seem to show any shift with the ASTE $^{12}$CO(3-2) map. They also compare the spatial distribution of CO with \spitz\ bands and found a fairly good correlation between the maps. As mentioned before, a peak in the 8 \mic\ emission, situated on the east side of N159S, corresponds to a peak in the H\,{\sc i} distribution while a peak in the 870 \mic\ emission (N159S) corresponds to a peak in the CO map. \citet{Fukui2009} already noted the offset between the H\,{\sc i} and CO peak for N159. They showed that the H\,{\sc i} and CO distributions correlate well on a 40-100 pc scale and argued that the H\,{\sc i} envelopes are gravitationally bound by giant molecular clouds. They also show that the correlation between the H\,{\sc i} and CO distribution breaks down on a more resolved scale. They interpret it as an illustration of the conversion of warmer, low-opacity H\,{\sc i} to colder high-opacity H\,{\sc i} from which H$_2$ could form, as suggested in \citet{Ott2008} and \citet{Wong2009}.

\subsubsection{Gas-to-dust mass ratios in N159}

Assuming a close mixing of dust and gas and that the gas-to-dust mass ratio does not depend on whether we study a predominantly atomic or molecular phase, we can relate the dust mass surface density in each resolved element of N159 to the gas reservoir such as: 
\begin{equation}
G/D=\frac{\Sigma_{HI}+X_{CO}I_{CO}}{\Sigma_{dust}}
\end{equation}

with X$_{CO}$ the conversion factor in units of \msun~pc$^{-2}$ (K km s$^{-1}$)$^{-1}$ and $\Sigma$$_{HI}$ and $\Sigma$$_{dust}$ the H\,{\sc i} and dust surface densities in \msun~pc$^{-2}$. \\

We first need to convert the ASTE CO(3-2) observation of the region N159 (shown in Fig.~\ref{LABOCA_HI_CO} bottom panel; private communication) to a CO(1-0) map. A mean CO (J = 3-2)/CO (J = 1-0) ratio (R$_{3-2/1-0}$) of 0.6 is usually derived in galaxy disks \citep{Warren2010} but R$_{3-2/1-0}$ can probe larger ranges \citep[0.2$<$R$_{3-2/1-0}$$<$2;][] {Meier2001,Tosaki2007,Mao2010,Warren2010}. \citet{Minamidani2008} investigate this ratio in 10 CO clumps of N159. This H\,{\sc ii} region being a hot and low-metallicity environment, they find values ranging between 0.7 and 1.4 (so higher than the median 0.6 value), with R$_{3-2/1-0}$=0.7 in the core of N159S and 1.1 in the cores of N159W and N159E. No clear North-South trend is observed however. We use the average R$_{3-2/1-0}$ value of the 10 clumps ($\sim$0.9) to convert the CO(3-2) into CO(1-0). From the H\,{\sc i} map and the derived CO(1-0) map, we can: \\

\vspace{-7pt}
{\it Derive a gas-to-dust mass ratio map -} We use a constant X$_{CO}$ factor to convert the CO intensity in molecular hydrogen mass surface density. \citet{Fukui2009} has performed a survey of molecular clouds in the LMC and estimate X$_{CO}$ to be $\sim$7 $\times$ 10$^{20}$ H$_2$ cm$^{-2}$ (K km s$^{-1}$)$^{-1}$, so three times the Galactic value of $\sim$2.3 $\times$ 10$^{20}$ H$_2$ cm$^{-2}$ (K km s$^{-1}$)$^{-1}$. We use the dust mass map obtained with the ``AC model" as our dust reference map and convolve the CO and dust surface density maps to the resolution of the H\,{\sc i} map using Gaussian kernels. Figure~\ref{DGR} (left) shows the gas(H\,{\sc i}+CO)-to-dust mass ratio map obtained with those hypothesis, with a mean gas-to-dust mass ratio (G/D) value of 384 (with a non-negligible uncertainty given the 50$\%$ errors on dust masses). As a reference, the Galactic value of G/D is $\sim$157 \citep{Zubko2004} so the value we derived is 2.4 times the solar value. The total H\,{\sc i} mass contained in the colored ISM elements of Fig.~\ref{DGR} is 4.3 $\times$ 10$^{5}$ \msun\ while the molecular gas mass is 1.1 $\times$ 10$^{6}$ \msun\ (thus 2.6 times higher than the atomic mass). The total dust mass in these ISM elements is 4.3 $\times$ 10$^{3}$ \msun, leading to a globally derived average of 356, consistent with the value derived on a resolved basis in spite of the local variations. We note that given the higher contribution of the H$_2$ to the total gas budget, using a slightly different (but constant) X$_{CO}$ factor would simply translate into a crude scaling of Fig.~\ref{DGR} to higher or lower gas-to-dust mass ratio ranges but would barely affect the qualitative distribution of Fig.~\ref{DGR} (left). Finally, dust masses are 2.8 times higher when graphite is used to model carbon dust (``graphite model"), leading to G/D smaller by the same factor (G/D$\sim$127). Studies of \citet{Franco_Cox_1986}, \citet{Lisenfeld_Ferrara_1998} or \citet{James2002}, among others, show that the G/D is proportional to the metallicity of the galaxy. The G/D obtained when dust masses are obtained with graphite in the SED modelling are thus too low compared to those expected in metal-poor environments like the LMC. \\
 
 \vspace{-7pt}
{\it Try to constrain the X$_{CO}$ factor -} If we assume now that G/D is constant across the region, we can use the technique proposed by \citet{Sandstrom2012_2} in order to derive the X$_{CO}$ factor that minimises the scatter of the logarithm of the G/D across the region. This technique minimises the effects of outliers on the measured scatter. As suggested in \citet{Sandstrom2012_2}, we tabulate the X$_{CO}$ values from 0.1 to 100 \msun~pc$^{-2}$ (K km s$^{-1}$)$^{-1}$ (thus from 6.2 $\times$ 10$^{18}$ to 6.2 $\times$ 10$^{21}$ H$_2$ cm$^{-2}$ (K km s$^{-1}$)$^{-1}$) and estimate the scatter of the logarithm of the G/D for each value of X$_{CO}$. The scatter for a given X$_{CO}$ value is estimated using the IDL {\it biweight$\_$mean.pro} function. We finally take the X$_{CO}$ value that minimises the scatter (Fig~\ref{DGR} - right).
The minimization occurs for X$_{CO}$=8.6 \msun~pc$^{-2}$ (K km s$^{-1}$)$^{-1}$ (= 5.4 $\times$ 10$^{20}$ H$_2$ cm$^{-2}$ (K km s$^{-1}$)$^{-1}$), thus 2.3 times the Galactic conversion factor and 1.3 times lower than that derived by \citet{Fukui2009}. This is slightly higher than the factor derived by \citet{Pineda2009} using CO observations only (and the virial mass/density profile of CO clumps) for the molecular ridge or the dust-based value derived by \citet{Leroy2011} ($\sim$4 $\times$ 10$^{20}$ H$_2$ cm$^{-2}$ (K km s$^{-1}$)$^{-1}$) for the LMC using \spitz/SAGE data only. 

We caution the fact that \citet{Sandstrom2012_2} apply the minimization method to unresolved molecular clouds. X$_{CO}$ is, in that case, a statistical quantity involving a cloud ensemble, namely a measure of the covering factor of CO clouds in more extended molecular structures. This method favors CO-bright gas and ignores CO-dark gas, which could naturally lead to X$_{CO}$ values close to those of the Solar Neighbourhood. This technique may not be applicable to the individual molecular clouds or clouds complexes we resolve in the LMC. 
Further studies on the X$_{CO}$ factor and potential CO-dark gas reservoirs using dust masses as a tracer of the gas masses will be performed in a future study.

%Because we expect the radiative environments of N159E/W and N159S to be different, we also carry out the same procedure on the two half-maps (region containing N159E/W or region containing N159S) separately. We find X$_{CO}$=4.7 \msun~pc$^{-2}$ (= 2.9 $\times$ 10$^{20}$ H$_2$ cm$^{-2}$ (K km s$^{-1}$)$^{-1}$) for N159E/W and X$_{CO}$=6.9 \msun~pc$^{-2}$ (= 4.3 $\times$ 10$^{20}$ H$_2$ cm$^{-2}$ (K km s$^{-1}$)$^{-1}$) for N159S. \\

%===============================================================

\section{Thermal emission at 870 \mic}

As mentioned previously, the LABOCA data reduction might remove faint diffuse emission in the outskirts of the complex. In this section, we select 24 regions across the complex where the 870 \mic\ emission is bright enough not to be affected by the filtering steps and analyse the thermal dust emission at 870 \mic\ in those regions. 
 
\subsection{Selected regions and correction of radio contamination}

\begin{figure*}
    \centering
    \begin{tabular}{ cc}
         \includegraphics[width=8cm ]{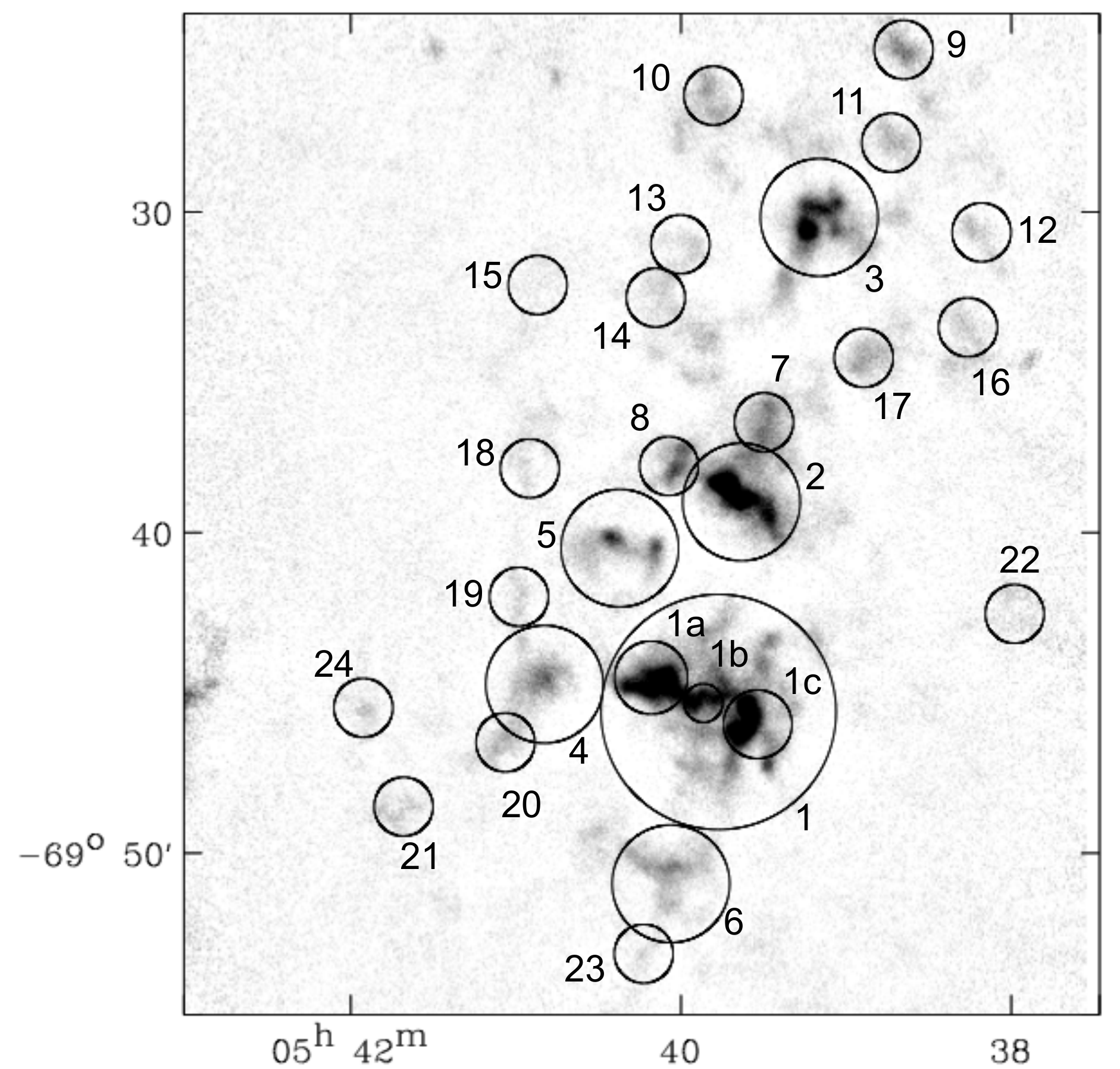}   &
         \includegraphics[width=8cm ]{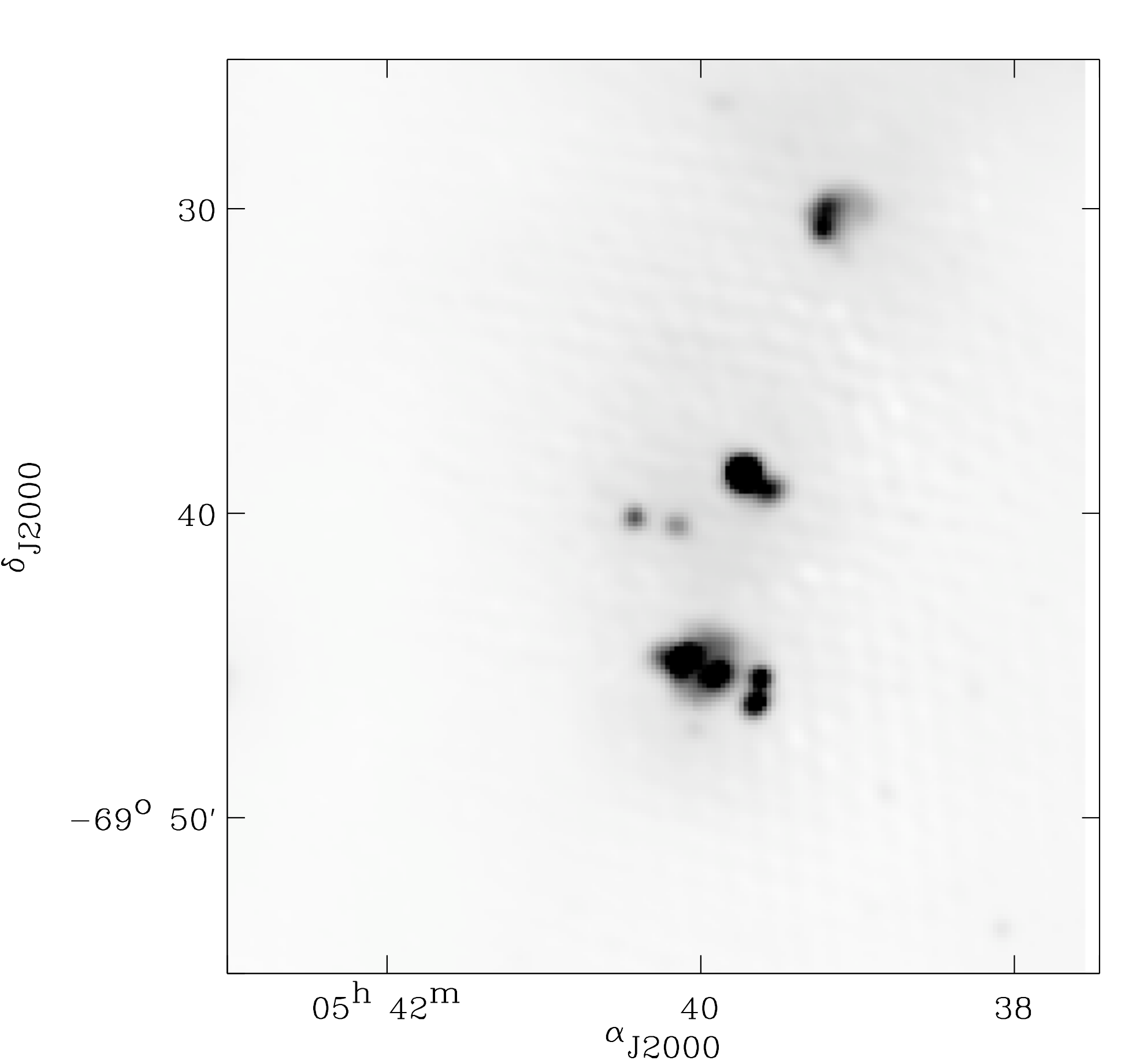} \\
         \end{tabular}
     \begin{tabular}{ c}
     \includegraphics[width=14cm ]{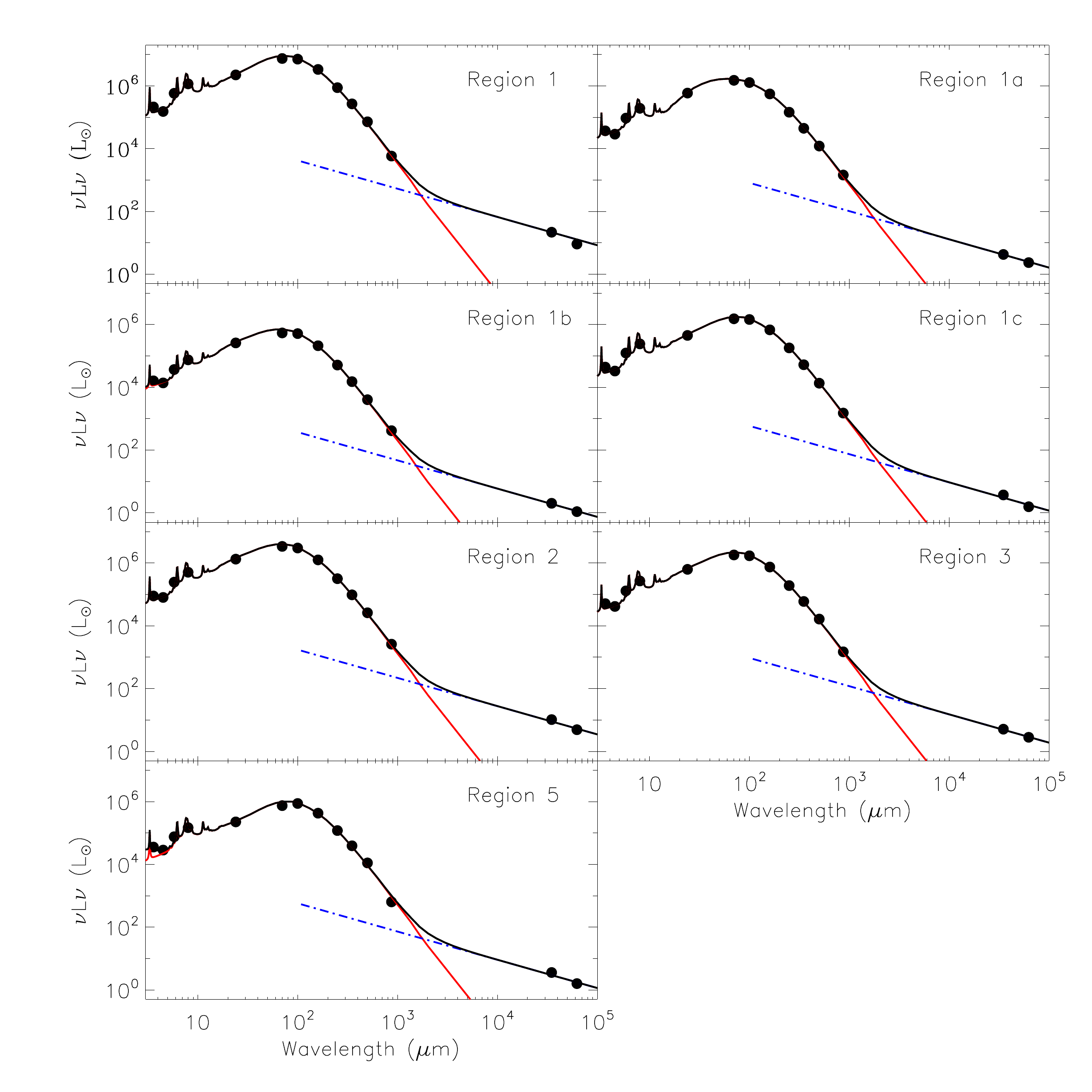}\\
        \end{tabular}
    \caption{{\it Top left panel:} Selected regions across the complex overlaid on the 870 \mic\ image. Circles also indicate the size of the photometric apertures. {\it Top right panel:} ATCA radio map at 4.8 GHz \citep[see][for the complete map of the LMC]{Dickel2005}. {\it Bottom panel:} Individual SEDs of the bright sources using the \citet{Galliano2011} ``AC" SED modelling technique in addition with a radio component. The SEDs models shown in this figure were thus produced using amorphous carbon to model carbon dust. Filled circles show \spitz, \hersc, \lab\ and ATCA radio fluxes. Errors are smaller than the symbols. The integrated SEDs appear with thick black lines. The red lines underline the thermal dust contribution. The dash-dotted lines indicate the free-free emission.}
    \label{SED_main}
\end{figure*}

We select individual regions across the complex that span a wide variety of environments, from intensely star-forming regions to a more quiescent ISM to study the detailed dust properties with the empirical SED model of \citet{Galliano2011}. Figure~\ref{SED_main} (top left) shows those regions. The circles indicate the photometric apertures we use. We detail the characteristics (centers and sizes) of those apertures in Table~\ref{Regions_Fluxes} along with the \hersc\ and LABOCA 870 \mic\ flux densities of the individual regions. \hersc\ errors are estimated using a combination in quadrature of calibration uncertainties and measurement uncertainties (background effects, choice of the aperture). \lab\ errors are quantified from the rms maps produced by the reduction pipeline. 

The 870 \mic\ flux densities we quote in Table~\ref{Regions_Fluxes} contain non-dust contribution, especially free-free radio emission. 4.8 (resolution 30\arcsec) and 8.6 GHz (resolution 15\arcsec) images of the entire LMC have been obtained by \citet{Dickel2005} using ATCA, which allow us to quantify the radio contamination. These observations reveal a strong radio emission in 30 Dor (peak of the radio emission of the entire LMC) as well as in the 3 bright star-forming regions N158, N159 and N160 (see Fig.~\ref{SED_main}, top right). They also indicate radio emission in two compact knots in the south east of N160, spatially correlated with peaks in the IR emission (as well as peaks in the SFR or $<$U$>$ maps) and corresponding to local decreases of the PAH fraction (see Fig.~\ref{ModelResults}).
We convolve the ATCA maps to the SPIRE 500 \mic\ resolution using a Gaussian kernel and estimate the 4.8 and 8.6 GHz fluxes (at 6.25 and 3.5 cm respectively) within our apertures for the 7 regions showing radio emission, namely region 1, 1a, 1b, 1c, 2, 3 and 5. With these two constraints, we perform a regression analysis (L$_{radio}$ $\propto$ $\nu$$^{-0.1}$) and estimate a free-free contamination of $\sim$9.6 $\%$, 9.0 $\%$ and 13 $\%$ of the 870 \mic\ fluxes in the regions 2, 3, 5 respectively and 10.4 $\%$ in region 1 (N159), with individual contaminations of 8.0 $\%$, 13 $\%$ and 5.6 $\%$ in the subregions 1a, 1b and 1c respectively.\\

%------------------------------------------------------------------------------------------------------------------------------------------
%------------------------------------------------------------------------------------------------------------------------------------------

 %----------------------------------------------------------------------------------------------------------------------------
\begin{figure*}
    \centering
\includegraphics[width=15cm]{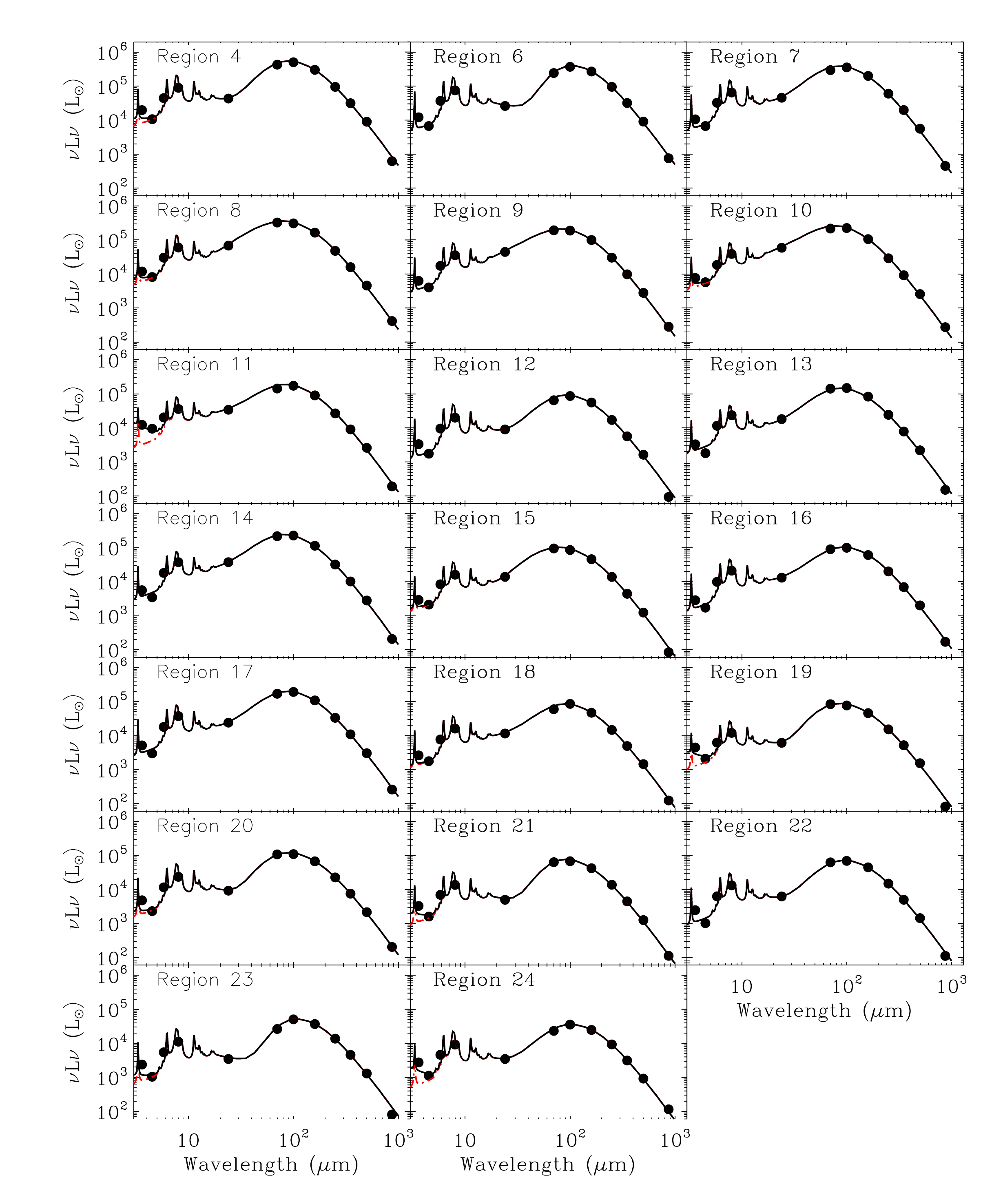}
    \caption{Individual SEDs in different regions of the diffuse ISM using the \citet{Galliano2011} ``AC" SED modelling technique. The SEDs models shown in this figure were thus produced using amorphous carbon to model carbon dust. Filled circles show \spitz, \hersc\ and \lab\ fluxes. Errors are smaller than the symbols.}
    \label{SED_second}
\end{figure*}
 %----------------------------------------------------------------------------------------------------------------------------

\subsection{Results}

We first apply our two-MBB model ($\beta$$_c$=1.5) to derive the cold dust temperature of our selected regions using the integrated flux densities tabulated in Table~\ref{Regions_Fluxes} but correcting the 870 \mic\ flux for radio contamination. We also model the selected regions using the "AC" model of \citet{Galliano2011} described in Section 3. For the 7 regions where radio emission is detected, we kept the 870 \mic\ flux densities of Table~\ref{Regions_Fluxes} (i.e. flux densities including non-dust contamination) but include the radio component in the modelling procedure to allow the visualization of this contribution in the SEDs of those regions. This component will be constrained by our 2 ATCA data points. As mentioned in the previous section, the free-free contamination is inferior to 10$\%$ in the H\,{\sc ii} region and negligible outside their cores.

Figure~\ref{SED_main} (bottom panel) shows the SEDs of regions detected in radio (and modelled with a free-free component) while Figure~\ref{SED_second} shows the SED obtained in the other regions. We observe variations in the SED shapes from region to region. The SEDs of the brightest H\,{\sc ii} regions (regions 1, 2, 3, 5) show `flat' profiles indicating a warmer dust temperature range. The main differences in the SED shape occur in the mid-infrared regime where the 24-to-70 slope strongly varies from one region to another, with higher 24/70 colour in H\,{\sc ii} regions and lower ratios in quiescent ISM (regions 12, 19 or 21 for instance). Region 6 (N159S) and region 21 associated with the same molecular cloud have SEDs strongly peaked due to the absence of star formation in the region.

Table~\ref{SED_model_results} summarises the most important parameters obtained from the two fitting procedures: the cold temperature T$_{dust}$, the dust mass M$_{dust}$, the PAH fraction f$_{PAH}$ and the three parameters characterising the radiation fields (the minimum and maximum intensities of the radiation field U$_{min}$ and U$_{max}$ and $\alpha$, the index describing the fraction of dust exposed to a given intensity). We also add the dust masses obtained when a ``graphite model" is applied to the data, for comparison.
As mentioned previously, diffuse ISM shows dust temperatures of $\sim$26K on average while N158, N159 and N160 show warmer dust temperatures ($>$30K). There is thus a clear evolution of the dust temperature range between bright star-forming regions like N159 (region 1), intermediate star-forming regions like region 3, 17, 16 to very quiescent regions like region 10 or 21. 
The dust masses are systematically lower when using amorphous carbon in lieu of graphite to model carbon dust, 2.8 times lower on average. Indeed, the ``amorphous carbon" model requires less cold dust to account for the same submm emission. This will have significant consequences on the study of the gas-to-dust mass ratio in the star-forming complex (see Section 6 for discussion). We finally note that the values derived using graphite are of the same order than those derived by \citet{Rantakyro2005} using SIMBA data at 1.2mm. 

The $\alpha$ values for the diffuse regions very often reach the limit of 2.5 fixed in the model for this parameter. This is consistent with the assumption that in the diffuse medium the heating intensity falls as r$^{-2}$ away from the heating source so dU/dr evolves as r$^{-3}$. In a uniform medium where dM$_{dust}$/dr evolves as r$^2$, dM$_{dust}$/dU is thus proportional to U$^{-2.5}$, so $\alpha$=2.5 \citep{Dale2001}. To perform a global SED modelling of nearby spiral galaxies, \citet{Draine2007} and \citet{Dale2012} advise fixing the maximum heating intensity to 10$^6$, to limit the number of free parameters \citep[fixed to 10$^{7}$ in][]{Aniano2012}. Given the number of data points we have to perform the modelling, we decide not to fix the maximum heating intensity of the radiation field U$_{max}$ in our study to investigate its possible variations from one region to another. Our U$_{max}$ values are not allowed to be higher than 10$^{7}$ during the modelling. Only two regions reach this maximum limit. N158, N159 and N160 show maximum intensities of $\sim$10$^4$-10$^5$ while diffuse ISM shows lower maximum intensities. As suggested by their peaked SED shape, the two south regions N159S and Region 23 have a very narrow (and low) range of radiation field intensities, with [U$_{min}$, U$_{max}$] ranges of [3.8-4.8] for Region 6, [3.0-4.0] for Region 23. As observed previously, the PAH fraction is very low in the H\,{\sc ii} regions compared to the diffuse ISM around those regions. 

Finally, we also estimate the H\,{\sc i} mass in each aperture. Values are tabulated in Table~\ref{SED_model_results}. As mentioned previously, the value of the Galactic G/D is 157. The LMC having a lower metallicity, the G/D of the LMC should be higher than this value. We observe a large spread in the H\,{\sc i}/D values from one region to another, and low ratios in H\,{\sc ii} regions, which is not surprising. Indeed, the whole region being a very large molecular complex, the molecular gas is thus expected to be a significant fraction of the total gas mass in those regions as well. As previously mentioned, further investigations including CO observations for the whole complex will be investigated in a future paper.

\subsection{SPIRE/LABOCA colour-colour diagram }

\begin{figure}
    \centering
\hspace{-25pt}\includegraphics[width=9.2cm]{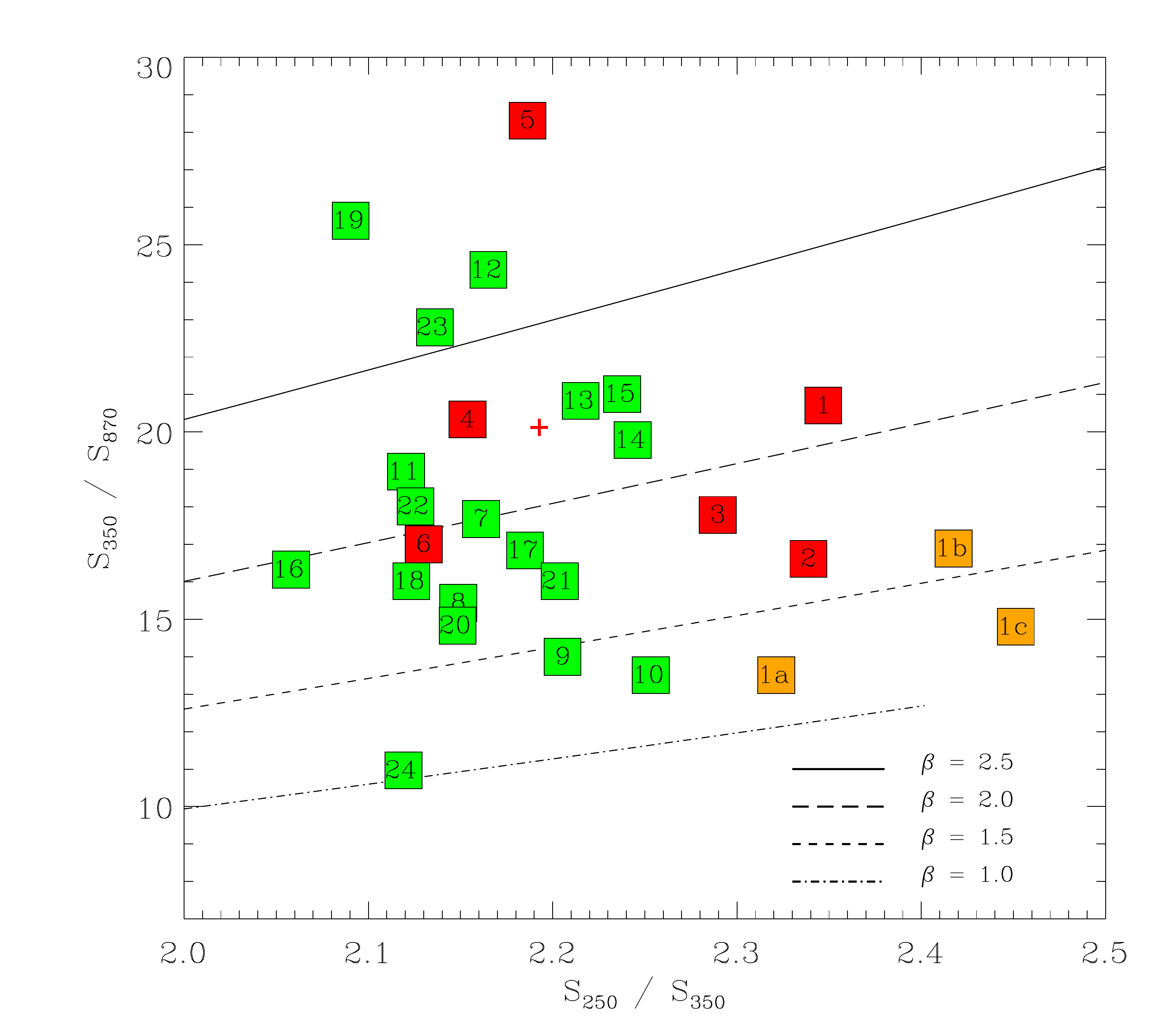}
    \caption{SPIRE/LABOCA colour-colour diagram. The 350-to-870 \mic\ flux density ratio are plotted as a function of the 250-to-350 \mic\ flux density ratio. The brightest star-forming regions of the complex (1 to 6) appear in red. The individual subregions of N159 (1a, 1b and 1c) are in orange. Regions 7 to 24 are in green. The lines show the colours obtained with single temperature modified blackbodies with an emissivity index of 2.5, 2, 1.5 and 1. Radio contamination has been removed from the 870 \mic\ flux densities. The red cross indicates where region 5 would fall if the modelled 870 \mic\ value  was used instead of the flux density we estimate.}
    \label{SPIRE_LABOCA_colour}
\end{figure}

Figure~\ref{SPIRE_LABOCA_colour} shows S$_{350}$/S$_{870}$ versus S$_{250}$/S$_{350}$. We remove the radio contamination from the LABOCA flux densities used in this plot. The different lines are colours obtained modelling single temperature modified blackbodies with an emissivity index of 2.5, 2, 1.5 and 1 and generated with temperatures ranging from 5 (left) to 70K (right). The bright star-forming regions have higher S$_{250}$/S$_{350}$, consistent with regions with higher temperature dust. They also show lower S$_{350}$/S$_{870}$ ratios. N158 (region 3), N159 (region 1) and N160 (region 2) thus present ``flatter" submm slope than regions of the diffuse ISM (i.e. higher 870 \mic\ flux density for a fixed 350 \mic\ flux density). Region 5 does not appear in the diagram because of its very high 350-to-870 colour. Figure~\ref{SED_main} (bottom panel) suggests that the 870 \mic\ flux density we derive could be underestimated compared to what is expected from the SPIRE bands. In Fig.~\ref{SPIRE_LABOCA_colour}, we indicate by a red cross where the region would fall in the plot if the modelled 870 \mic\ value (derived from our fit) was used instead of the flux density derived directly from the 870 \mic\ map. 

 %----------------------------------------------------------------------------------------------------------------------------
 %----------------------------------------------------------------------------------------------------------------------------

\subsection{About submm excess}

Ground-based data and now Herschel observations of low-metallicity environments at submm wavelengths have helped us to better probe the cold phases of dust and investigate variations in their properties. These observations have led to the detection of a ``submm excess" \citep{Galliano2005,Marleau2006,Galametz2009}, namely submm emission higher than that predicted by FIR data and standard dust models \citep[][for instance]{Draine_Li_2007}. Several interpretations have been analysed to explain this excess. It could be linked with a different dust component:
very cold dust \citep[e.g.][]{Galliano2003,Galliano2005,Galametz2010}, emission from ``spinning dust"  \citep{Ysard2010} or magnetic dipole radiation from magnetic nano-particles \citep{Draine2012}. The excess could also be linked with a modification of the dust emissivity properties at colder temperatures \citep{Meny2007,Paradis2010}.

Several studies have been dedicated to the investigation of submm and millimeter excess in the LMC. \citet{Bot2010} found excess millimeter emission while modelling the integrated SED of the LMC using FIR and radio data. This excess was explained by possible emission from ``spinning dust". However, this hypothesis could require extreme excitation conditions for the small dust grains expected to be the driver of such emission. Recent {\it Planck} observations of the LMC were analysed in \citet{Planck_collabo_2011_MagellanicClouds} and also lead to the detection of a millimeter excess longward 500 \mic. They found that CMB fluctuations would be sufficient to explain the excess.
Using \hersc\ observations of a large strip across the LMC, \citet{Gordon2010} showed, on a more resolved scale, that the \hersc\ 500 \mic\ emission was $\sim$10$\%$ higher than expected from extrapolating models fitted at shorter wavelengths (and using a modified blackbody with an emissivity index of $\beta$=1.5) and found a weak anti-correlation of the excess with the MIPS 24 \mic\ emission or the total gas surface density. Using the resolved SED modelling technique we use in this paper (and AC to model carbon dust), \citet{Galliano2011} confirm the detection of a 500 \mic\ extended emission excess in the same strip across the LMC, with an average relative amplitude of 15 $\%$ (and variations up to 40 $\%$). They found that this resolved submm excess anti-correlates well with the dust mass surface density and probably not linked with very cold dust or CMB fluctuations.

The SEDs presented in Fig.~\ref{SED_main} and Fig.~\ref{SED_second} do not show excess emission long-ward 500 \mic. Our model seems sufficient to properly fit the observations in the strongly star-forming regions as well as in the more diffuse regions we modelled, without requiring any extra component to explain the submm emission. We nevertheless note that unlike \citet{Gordon2010} or \citet{Galliano2011}, we do not study regions with very low surface densities in this analysis. If the environment where submm excess is hiding is in the low surface brightness regions, we will not be able to detect it in our star-forming complex. Low surface brightness structures are also the regions where our data reduction of LABOCA data is the most likely to have removed signal, which could explain the non-detection.

 %===============================================================
% Conclusion and summary
%===============================================================

\section{Conclusions}

In this paper, we combine \spitz\ Space Telescope, \hersc\ Space Observatory and LABOCA mid-infrared to submm data (from 3.6 to 870 \mic) of the resolved massive star-forming region N158-N159-N160 located in the Large Magellanic Cloud. 

\begin{itemize}

  \item Using a resolved 2MBB fitting technique and a $\beta$$_c$ fixed to 1.5, we find an average temperature of $\sim$27K across the region, with a maximum in N160. Using data from 3.6 to 500 \mic, a physical SED modelling and the hypothesis of amorphous carbon to model carbon dust, we also derive maps of the dust temperature distribution, the average starlight intensity, the PAH fraction, the star formation rate and the dust mass surface densities. The PAH fraction strongly decreases in our H\,{\sc ii} regions. This decrease coincides with peaks in the mean radiation field intensity map, which is consistent with the well-known destruction of PAHs by strong radiation fields. The absence of PAHs could also be linked with the young age of the clusters, namely that the regions could be too young to have formed any PAH.

  \item We analyse how submm colours vary across the star-forming complex and study their dependency with the surface brightnesses or the radiation field intensity $<$U$>$. We find that the submm ratios correlate strongly with $<$U$>$ and that while the 100/160 and 160/250 colours correlate more strongly with the 24 \mic\ brightness, the 250/350 and 350/500 colours equally correlate with the 3.6 and the 24 \mic\ surface brightnesses. Possible variation of the emissivity with temperature have been investigated using SPIRE colour-colour diagrams. No clear trend is observed.
  
  \item The dust surface densities follow the FIR distribution and a non-negligible amount of dust is detected in N159S. We find a total dust mass of 2.1 $\times$ 10$^4$ \msun\ in the resolved elements we model (and 2.8 times more dust if graphite is used to model carbon dust). 
  
  \item Considering that gas and dust are closely mixed in the N159 region, we use our dust mass map to compare the distribution of dust with those of the tracers of the atomic gas in the complex to investigate variations in the atomic+molecular gas-to-dust ratio in N159. We use a CO(3-2) map of N159 to quantify the molecular reservoir in that region. A mean value of $\sim$356 is derived for the N159 complex when using the X$_{CO}$ conversion factor of \citet{Fukui2009} (=7 $\times$ 10$^{20}$ H$_2$ cm$^{-2}$ (K km s$^{-1}$)$^{-1}$).  If we consider a constant G/D in the complex, we can apply the minimization technique of \citet{Sandstrom2012_2} to inversely derive the X$_{CO}$ conversion factor of the complex that minimises the D/G scatter in N159. The D/G scatter is minimized when the conversion factor X$_{CO}$ is equal to 8.6 \msun~pc$^{-2}$ (K km s$^{-1}$)$^{-1}$ (=5.4$\times$10$^{20}$ H$_2$ cm$^{-2}$ (K km s$^{-1}$)$^{-1}$).

   \item We finally model the SEDs of 24 selected regions, now including the 870 \mic\ data in the fitting, and describe the variations in the dust thermal emission (thus in the SED shape) across the complex. We show that our ``AC" model is sufficient to explain the submm emission we observe at 870 \mic.

\end{itemize}

%===============================================================
% Acknowledgements
%===============================================================

\section*{Acknowledgments}
First, we would like to thank the referee for his/her careful reading of the paper. M.R. wishes to acknowledge support from FONDECYT(CHILE) grant No1080335. PACS has been developed by MPE (Germany); UVIE (Austria); KU Leuven, CSL, IMEC (Belgium); CEA, LAM (France); MPIA (Germany); INAF-IFSI/OAA/OAP/OAT, LENS, SISSA (Italy); IAC (Spain). This development has been supported by BMVIT (Austria), ESA-PRODEX (Belgium), CEA/CNES (France), DLR (Germany), ASI/INAF (Italy), and CICYT/MCYT (Spain). 
SPIRE has been developed by a consortium of institutes led by Cardiff Univ. (UK) and including: Univ. Lethbridge (Canada); NAOC (China); CEA, LAM (France); IFSI, Univ. Padua (Italy);IAC (Spain); Stockholm Observatory (Sweden); Imperial College London, RAL, UCL-MSSL, UKATC, Univ. Sussex (UK); and Caltech, JPL, NHSC, Univ. Colorado (USA). This development has been supported by national funding agencies: CSA (Canada); NAOC (China); CEA, CNES, CNRS (France); ASI (Italy); MCINN (Spain); SNSB (Sweden); STFC, UKSA (UK); and NASA (USA).

%===============================================================
% Bibliography
%===============================================================

\bibliographystyle{mn2e}
\bibliography{/Users/maudgalametz/Documents/Work/Papers/mybiblio.bib}

%---------------------------- Regions_Fluxes -------------------------------------------------------------------------------------------
\begin{landscape}
\begin{table}
\caption{MIPS, PACS, SPIRE and LABOCA flux densities}
\label{Regions_Fluxes}
 \centering
 \begin{tabular}{ccccccccccccc}
\hline
\hline
&&&&&&&&&&&&\\
Region &     \multicolumn{2}{c}{Center} & Radius  & MIPS 24 & MIPS 70 & PACS 100 &  PACS 160 & SPIRE 250 & SPIRE 350 & SPIRE 500 & LABOCA $^a$ \\
              & RA (J2000) & DEC (J2000)    & (arcsec) & (Jy) & (Jy)  & (Jy)              &  (Jy)                  &  (Jy)             &  (Jy)             &  (Jy) &  (Jy)  \\
              &&&&&&&&&&&&\\
     \hline
     &&&&&&&&&&&&\\
1 (N159)   & 05 39 46.65 & -69 45 39.94 		& 220 & 243.5$\pm$24.3 & 2363.5$\pm$236.3 & 3226.9$\pm$163.5& 2427.2$\pm$121.4 & 995.4$\pm$70.2	& 424.2$\pm$29.9	& 162.5$\pm$11.5	& 22.9$\pm$1.6 \\
1a (N159E) & 05 40 10.98 & -69 44 35.84 	& 64   &  64.3$\pm$6.4     & 477.9$\pm$47.8    	& 579.5$\pm$29.1 	& 402.0$\pm$20.1 	& 164.1$\pm$11.6	& 70.7$\pm$5.0	& 27.5$\pm$2.0	& 5.7$\pm$0.1 \\
1b & 05 39 52.10 & -69 45 23.17 			& 36   &    28.1$\pm$2.8   & 170.3$\pm$17.0 	& 231.8$\pm$11.6 	& 152.4$\pm$7.6 	& 57.7$\pm$4.1	& 23.9$\pm$1.7	& 9.1$\pm$0.6		& 1.6$\pm$0.1 \\
1c (N159W) & 05 39 32.51 & -69 46 02.74 	& 68  & 48.7$\pm$4.9       & 481.2$\pm$48.1 	& 651.8$\pm$32.7 	& 487.7$\pm$24.4 	& 202.6$\pm$14.4	& 82.6$\pm$5.9	& 30.4$\pm$2.2	& 5.9$\pm$0.2 \\
2  (N160)   & 05 39 38.63 & -69 39 06.79 	         & 110 & 144.7$\pm$14.5 & 1073.3$\pm$107.3	& 1362.3$\pm$68.5 & 912.9$\pm$45.6 	& 359.2$\pm$25.5	& 153.6$\pm$10.9   & 58.9$\pm$4.2		& 10.2$\pm$0.4 \\
3  (N158) & 05 39 11.22 & -69 30 13.65 		& 110 &  68.3$\pm$6.8    & 572.9$\pm$57.3 	& 770.2$\pm$39.1 	& 540.8$\pm$27.0 	& 216.2$\pm$15.3	& 94.4$\pm$6.7	& 37.1$\pm$2.6	& 5.8$\pm$0.4 \\
4   & 05 40 49.41 & -69 44 48.22 			& 110 &  4.7$\pm$0.5       & 134.5$\pm$13.4 	& 226.2$\pm$12.5	 & 217.0$\pm$10.9 	& 106.88$\pm$7.5	& 49.6$\pm$3.5	& 20.2$\pm$1.4	& 2.4$\pm$0.4 \\
5   & 05 40 22.25 & -69 40 33.51 			& 110 &  24.5$\pm$2.4     & 235.6$\pm$23.6 	& 394.6$\pm$20.6	 & 309.6$\pm$15.5 	& 135.2$\pm$9.6	& 61.9$\pm$4.4	& 25.2$\pm$1.8	& 2.5$\pm$0.4 \\
6 (N159S)   & 05 40 03.75 & -69 51 01.62 	& 110 &  2.8$\pm$0.3      & 76.3$\pm$7.6 		& 166.2$\pm$9.6 	& 193.9$\pm$9.7 	& 107.3$\pm$7.5	& 50.4$\pm$3.5	& 20.4$\pm$1.4	& 3.0$\pm$0.4 \\
7 & 05 39 30.65 & -69 36 37.42 			& 55  &    4.9$\pm$0.5     & 93.1$\pm$9.3 		& 159.6$\pm$8.2 	& 143.8$\pm$7.2 	& 67.4$\pm$4.7	& 31.2$\pm$2.2	& 12.6$\pm$0.9	& 1.8$\pm$0.1 \\
8 & 05 40 04.63 & -69 37 59.86 			& 55  &     7.4$\pm$0.7 & 102.6$\pm$10.3 	& 138.6$\pm$7.1 	& 118.2$\pm$5.9 	& 53.9$\pm$3.8	& 25.1$\pm$1.8	& 10.3$\pm$0.7	& 1.6$\pm$0.1 \\
9   & 05 38 41.55 & -69 24 58.82 			& 55  &     4.8$\pm$0.5 & 60.7$\pm$6.1 		& 84.8$\pm$4.5 	& 71.9$\pm$3.6 	& 34.0$\pm$2.4	& 15.4$\pm$1.1	& 6.3$\pm$0.5		& 1.1$\pm$0.1 \\
10   & 05 39 49.03 & -69 26 26.38 			& 55  &        6.4$\pm$0.6 & 68.2$\pm$6.8 		& 100.9$\pm$5.3	 & 76.4$\pm$3.8 	& 32.6$\pm$2.3	& 14.5$\pm$1.0	& 5.8$\pm$0.4		& 1.1$\pm$0.1 \\
11  & 05 38 45.73 & -69 27 53.10 			& 55  &     3.7$\pm$0.4 & 45.2$\pm$4.5 		& 78.0$\pm$4.2 	& 65.0$\pm$3.3 	& 30.1$\pm$2.1	& 14.2$\pm$1.0	& 5.9$\pm$0.4		& 0.8$\pm$0.1 \\
12 & 05 38 13.31 & -69 30 39.46 			& 55  &       1.0$\pm$0.1 & 20.5$\pm$2.0 		& 38.9$\pm$2.3 	& 40.2$\pm$2.0 	& 19.2$\pm$1.4	& 8.9$\pm$0.6		& 3.7$\pm$0.3		& 0.4$\pm$0.1 \\
13 & 05 40 00.70 & -69 31 04.85 			& 55  &     2.0$\pm$0.2 & 44.9$\pm$4.5 		& 66.8$\pm$3.6 	& 59.6$\pm$3.0 	& 27.3$\pm$1.9	& 12.3$\pm$0.9	& 4.9$\pm$0.4		& 0.6$\pm$0.1 \\
14 & 05 40 09.44 & -69 32 44.60			& 55  &     4.1$\pm$0.4 & 68.8$\pm$6.9 		& 103.0$\pm$5.4 	& 82.6$\pm$4.1 	& 36.2$\pm$2.5	& 16.1$\pm$1.1	& 6.3$\pm$0.5		& 0.8$\pm$0.1 \\
15 & 05 40 51.61 & -69 32 20.98 			& 55  &     1.5$\pm$0.1 & 30.1$\pm$3.0 		& 38.6$\pm$2.3 	& 33.2$\pm$1.7 	& 15.7$\pm$1.1	& 7.0$\pm$0.5		& 2.8$\pm$0.2		& 0.3$\pm$0.1 \\
16 & 05 38 17.88 & -69 33 37.52 			& 55  &      1.4$\pm$0.1 & 28.4$\pm$2.8 		& 45.0$\pm$2.5 	& 43.9$\pm$2.2 	& 22.6$\pm$1.6	& 11.0$\pm$0.8	& 4.5$\pm$0.3		& 0.7$\pm$0.1 \\
17 & 05 38 55.03 & -69 34 36.05 			& 55  &      2.6$\pm$0.3 & 53.8$\pm$5.4 		& 86.8$\pm$4.6 	& 78.3$\pm$3.9	& 37.6$\pm$2.6	& 17.2$\pm$1.2	& 6.8$\pm$0.5		& 1.0$\pm$0.1 \\
18 & 05 40 54.56 & -69 38 04.19 			& 55  &     1.3$\pm$0.1 & 18.8$\pm$1.9 		& 38.6$\pm$2.4 	& 34.0$\pm$1.7 	& 16.6$\pm$1.2	& 7.8$\pm$0.6		& 3.3$\pm$0.2		& 0.5$\pm$0.1 \\
19 & 05 40 58.54 & -69 42 04.22 			& 55  &      0.7$\pm$0.1 & 26.5$\pm$2.6 		& 34.7$\pm$2.1 	& 33.1$\pm$1.7 	& 17.2$\pm$1.2	& 8.2$\pm$0.6		& 3.5$\pm$0.3		& 0.3$\pm$0.1 \\
20 & 05 41 03.54 & -69 46 36.87 			& 55  &    1.0$\pm$0.1 & 33.9$\pm$3.4 		& 49.3$\pm$2.8 	& 48.5$\pm$2.4 	& 25.5$\pm$1.8	& 11.9$\pm$0.8	& 4.9$\pm$0.3		& 0.8$\pm$0.1 \\
21 & 05 41 40.44 & -69 48 36.23 			& 55  &      0.5$\pm$0.1 & 20.0$\pm$2.0 		& 30.6$\pm$2.3 	& 30.4$\pm$1.5 	& 15.5$\pm$1.1	& 7.1$\pm$0.5		& 2.9$\pm$0.2		& 0.4$\pm$0.1 \\
22 & 05 38 00.17 & -69 42 32.98 			& 55  &    0.7$\pm$0.1 & 19.8$\pm$2.0 		& 31.2$\pm$1.9 	& 32.4$\pm$1.6 	& 16.8$\pm$1.2	& 7.9$\pm$0.6		& 3.2$\pm$0.2		& 0.4$\pm$0.1 \\
23 & 05 40 13.59 & -69 53 12.56 			& 55  &    0.4$\pm$0.1 & 8.34$\pm$0.8 		& 22.9$\pm$1.6 	& 26.7$\pm$1.3   	& 15.5$\pm$1.1	& 7.3$\pm$0.5		& 2.9$\pm$0.2		& 0.3$\pm$0.1 \\
24 & 05 41 54.74 & -69 45 30.83 			& 55  &   0.4$\pm$0.1 & 7.3$\pm$0.7 		& 16.1$\pm$1.8 	& 17.9$\pm$0.9 	& 10.5$\pm$0.8	& 5.0$\pm$0.4		& 2.1$\pm$0.2		& 0.5$\pm$0.1 \\

&&&&&&&&&&&&\\
 \hline
\end{tabular}
\begin{list}{}{}
\item[$^a$] Radio contamination is not subtracted from those values.
\end{list}
 \end{table} 
 \end{landscape}
 %----------------------------------------------------------------------------------------------------------------------------
%---------------------------- SED modelling results-------------------------------------------------------------------------------------------
\begin{table*}
\caption{Cold dust temperatures, dust masses, PAH fraction, radiation field intensity parameters and H\,{\sc i}-to-dust mass ratios obtained for our selected regions using the ``AC model"}
\label{SED_model_results}
 \centering
 \begin{tabular}{cccccccc}
\hline
\hline
 &\\
Region &   T$_{c}$ & M$_{dust}$ & f$_{PAH}$ & $\alpha$ & U$_{min}$ & U$_{max}$ & H\,{\sc i}/D \\
& (K) & (\msun) & \\
 &\\
     \hline  
 &\\
 1 (N159)  	& 34.7 &  3.08 $\times$ 10$^{3}$ (8.67 $\times$ 10$^{3}$)$^a$ &0.37	&2.38	&6.9		&1.2 $\times$ 10$^{5}$	& 165 \\
1a (N159E) 	& 27.7 &  8.01 $\times$ 10$^{2}$ (3.40 $\times$ 10$^{3}$)	&0.27	&1.94	&2.2		&3.5 $\times$ 10$^{3}$	& 55 \\
1b 			& 34.7 &  1.71 $\times$ 10$^{2}$ (6.03 $\times$ 10$^{2}$) 	&0.23	&2.26	&8.2		&1.0 $\times$ 10$^{5}$	& 79\\
1c (N159W) 	& 31.0 &  7.50 $\times$ 10$^{2}$ (2.92 $\times$ 10$^{3}$)	&0.41	&2.32	&5.1		&3.9 $\times$ 10$^{4}$	& 67 \\
2 (N160)  		& 33.7 &  1.15 $\times$ 10$^{3}$ (4.32 $\times$ 10$^{3}$)	&0.31	&2.27	&7.1		&1.1 $\times$ 10$^{5}$	& 102\\
3 (N158) 		& 33.6 &  7.21 $\times$ 10$^{2}$ (2.17 $\times$ 10$^{3}$)	&0.32	&2.31	&6.5		&9.4 $\times$ 10$^{4}$	& 130\\
4   			& 26.8 &  5.29 $\times$ 10$^{2}$ (1.44 $\times$ 10$^{3}$)	&0.63	&2.50	&2.8		&2.2 $\times$ 10$^{2}$	& 199 \\
5  			& 34.7 &  5.14 $\times$ 10$^{2}$ (1.36 $\times$ 10$^{3}$)	&0.39	&2.39	&4.6		&1.0 $\times$ 10$^{7}$	& 195\\
6 (N159S)  	& 24.8 &  6.44 $\times$ 10$^{2}$ (1.92 $\times$ 10$^{3}$)	&0.83	&2.42	&3.8		&4.8					& 212 \\
7  			& 28.7 &  3.01 $\times$ 10$^{2}$ (8.65 $\times$ 10$^{2}$)	&0.60	&2.50	&3.4		&3.1 $\times$ 10$^{3}$	& 99 \\
8  			& 27.2 &  2.65 $\times$ 10$^{2}$ (9.18 $\times$ 10$^{2}$)	&0.54	&2.24	&2.7		&3.9 $\times$ 10$^{3}$	& 131 \\
9      			& 26.7 &  1.66 $\times$ 10$^{2}$ (4.91 $\times$ 10$^{2}$)	&0.53	&2.32	&2.8		&2.1 $\times$ 10$^{3}$	& 111 \\
10   			& 29.7 &  1.34 $\times$ 10$^{2}$ (4.04 $\times$ 10$^{2}$)	&0.44	&2.36	&4.4		&1.3 $\times$ 10$^{5}$	& 126 \\
11   			& 28.5 &  1.42 $\times$ 10$^{2}$ (3.87 $\times$ 10$^{2}$)	&0.60	&2.43	&3.2		&1.0 $\times$ 10$^{7}$	& 151 \\
12   			& 26.8 &  1.08 $\times$ 10$^{2}$ (2.47 $\times$ 10$^{2}$)	&0.86	&2.50	&2.3		&5.5 $\times$ 10$^{2}$	& 164\\
13   			& 28.2 &  1.29 $\times$ 10$^{2}$ (3.10 $\times$ 10$^{2}$)	&0.48	&2.50	&2.9		&4.8 $\times$ 10$^{3}$	& 205 \\
14   			& 30.3 &  1.45 $\times$ 10$^{2}$ (3.66 $\times$ 10$^{2}$)	&0.47	&2.50	&4.3		&4.2 $\times$ 10$^{4}$	& 196 \\
15  			& 26.8 &  7.34 $\times$ 10$^{1}$ (2.16 $\times$ 10$^{2}$)	&0.54	&2.18	&2.7		&9.0 $\times$ 10$^{3}$	& 374 \\
16  			& 24.9 &  1.34 $\times$ 10$^{2}$ (3.57 $\times$ 10$^{2}$)	&0.67	&2.50	&2.0		&2.0 $\times$ 10$^{4}$	& 138 \\
17  			& 27.5 &  1.84 $\times$ 10$^{2}$ (4.60 $\times$ 10$^{2}$)	&0.62	&2.50	&2.8		&4.1 $\times$ 10$^{3}$	& 145\\
18  			& 27.2 &  9.20 $\times$ 10$^{1}$ (2.25 $\times$ 10$^{2}$)	&0.66	&2.50	&2.4		&8.2 $\times$ 10$^{4}$	& 219 \\
19   			& 24.2 &  1.09 $\times$ 10$^{2}$ (3.30 $\times$ 10$^{2}$)	&0.48	&1.75	&1.0		&7.3 $\times$ 10$^{1}$	& 202 \\
20   			& 24.5 &  1.64 $\times$ 10$^{2}$ (5.37 $\times$ 10$^{2}$)	&0.78	&1.78	&1.1		&5.5 $\times$ 10$^{1}$	& 176\\
21   			& 25.6 &  8.79 $\times$ 10$^{1}$ (3.23 $\times$ 10$^{2}$)	&0.74	&1.00	&1.0		&2.1 $\times$ 10$^{1}$	& 327 \\
22   			&  24.8 & 9.97 $\times$ 10$^{1}$ (2.47 $\times$ 10$^{2}$)	&0.67	&2.50	&1.9		&3.5 $\times$ 10$^{2}$	& 275 \\
23   			&  24.1 & 1.07 $\times$ 10$^{2}$ (2.60 $\times$ 10$^{2}$)	&0.95	&2.50	&3.0		&4.0					& 289 \\
24   			& 23.9 &  7.76 $\times$ 10$^{1}$ (1.92 $\times$ 10$^{2}$)	&0.99	&2.50	&1.3		&4.6 	$\times$ 10$^{2}$	& 182 \\
 &\\
\hline
\end{tabular}
\begin{list}{}{}
\item[$^a$] We indicate the dust masses obtained using the graphite model in parenthesis for comparison. 
\end{list}
 \end{table*} 
 %----------------------------------------------------------------------------------------------------------------------------

\end{document}